\documentclass{aa}  
\usepackage{graphicx}
\usepackage{array}
\usepackage{lscape}                                
\usepackage{natbib}
\usepackage{longtable}
\usepackage{color}
\usepackage{mathrsfs} 
\usepackage{txfonts,textcomp}
\usepackage{tikz}
\usepackage{amssymb}
\usepackage{multirow}
\usepackage{pdflscape}
\usepackage{txfonts}

\bibpunct{(}{)}{;}{a}{}{,}

\begin{document}

\title{Combined asteroseismology, spectroscopy, and astrometry of 
the CoRoT B2V target HD\,170580\thanks{Based on
    photometric data assembles with the CoRoT space mission, which was developed
    and operated by the French space agency CNES, with participation of ESA's
    RSSD and Science Programmes, Austria, Belgium, Brazil, Germany and
    Spain. Based on spectroscopic observations made with the HERMES spectrograph
    supported by the Research Foundation Flanders (FWO), Belgium, the Research
    Council of KU\,Leuven, Belgium, the Fonds National Recherches Scientific
    (FNRS), Belgium, the Royal Observatory of Belgium, the Observatoire de
    Gen\`eve, Switzerland, and the Th\"{u}ringer Landessternwarte Tautenburg,
    Germany; HERMES is attached to the Mercator telescope, operated by the
    Flemish Community on the island of La Palma at the Spanish Observatorio del
    Roque de los Muchachos of the Instituto de Astrof\'{i}sica de Canarias. Also
    based on spectroscopy made with the ESO 3.6m telescope at La Silla
    Observatory under the ESO Large Programme LP182.D-0356. }}

\author{C. Aerts\inst{1,2,3}\and 
M. G. Pedersen\inst{1}\and
E. Vermeyen\inst{1}\and
L.\ Hendriks\inst{2}\and 
C.\ Johnston\inst{1}\and 
A.\ Tkachenko\inst{1}\and
P.\ I.\ P\'{a}pics\inst{1}\and 
J.\ Debosscher\inst{1}\and 
M.\ Briquet\inst{4}\and
A.\ Thoul\inst{3,4}\and 
M.\ Rainer\inst{5,6}\and 
E.\ Poretti\inst{6}}

\institute{Instituut voor Sterrenkunde, KU Leuven, Celestijnenlaan 200D, 3001
  Leuven, Belgium\\ \email{Conny.Aerts@ster.kuleuven.be} \and Department of
  Astrophysics/IMAPP, Radboud University Nijmegen, 6500 GL Nijmegen, The
  Netherlands \and Kavli Institute for Theoretical Physics, University of
  California, Santa Barbara, CA 93106, USA \and 
Space sciences, Technologies and Astrophysics Research (STAR) Institute, 
Universit\'e de Li\`ege, 17 All\'ee du 6 Ao\^ut 19C, Bat. B5C, 4000, Li\`ege,
  Belgium \and 
INAF - Osservatorio Astrofisico di Arcetri
Largo E. Fermi 5, 50125 Firenze, Italy
\and
INAF - Osservatorio Astronomico di Brera, via E.\ Bianchi 46, 23807
  Merate, Italy 
}

\offprints{Conny.Aerts@kuleuven.be}

\date{Submitted/Accepted}

\titlerunning{Asteroseismology of the CoRoT B2V star HD\,170580}
\authorrunning{C.\ Aerts et al.}

\abstract{Space asteroseismology reveals that stellar structure and evolution
  models of intermediate- and high-mass stars are in need of improvement in
  terms of angular momentum and chemical element transport.}  {We aim to probe
  the interior structure of a hot massive star in the core-hydrogen burning
  phase of its evolution.}  {We analyse CoRoT space photometry, Gaia DR2 space
  astrometry, and high-resolution high signal-to-noise HERMES and HARPS
  time-series spectroscopy of the slowly rotating B2V star HD\,170580.}  {From
  the time-series spectroscopy we derive $v\sin i=4\pm 2\,$km\,s$^{-1}$, where
  the uncertainty results from the complex pulsational line-profile variability
  that was so far ignored in the literature.  We detect { 42 frequencies with
    amplitude above five times the local noise level. Among these we identify
    five rotationally split triplets and one quintuplet.}  Asteroseismic
  modelling based on CoRoT, Gaia DR2 and spectroscopic data leads to a star of
  $M\sim 8\,$M$_\odot$ near core-hydrogen exhaustion and an extended overshoot
  zone.  The detected { low-order pressure-mode} frequencies cannot be fit
  within the uncertainties of the CoRoT data by models without atomic diffusion.
  Irrespective of this limitation, { the low-order gravity modes reveal
    HD\,170580 to be a slow rotator with an average rotation period between 
      73 and 98\,d and a hint of small differential rotation.}}  {Future Gaia
    DR3 data taking into account the multiplicity of { the star, along with
      long-term TESS photometry would allow} to put better observational
    constraints on the asteroseismic models of this blue evolved massive star.
    { Improved modelling with atomic diffusion, including radiative
      levitation, is needed to achieve compliance} with the low helium surface
    abundance of the star. This poses immense computational challenges { but
      is required to derive the interior rotation and mixing profiles of this
      star.}}


\maketitle

\section{Introduction}

In the context of the chemical evolution of galaxies, stars of spectal type O
and B play a dominant role.
It is therefore essential to know the amount of helium in the core at the end of
their main-sequence phase (TAMS), because this determines the efficiency and
amount of heavy element production during the final 10\% of such stars'
lives. Determining the helium core mass of OB-type stars, which are born with a
well-developed convective core and a radiative envelope, requires a quantitative
estimate of the level and shape of the core overshooting properties and chemical
mixing in the envelope for a representative sample of such stars. This sample
must cover the entire main sequence and various levels of stellar rotation.
Traditionally, the macroscopic envelope mixing in radiatively stratified layers
of stars has been assumed to have a rotational origin
\citep[e.g.,][]{Brott2011}. However, this does not offer a satisfactory
explanation for slowly rotating OB stars that are nitrogen enriched
\citep[e.g.][]{Dufton2018}. Moreover, pulsational mixing was found to be an
alternative interpretation for slow to moderate rotators based on observational
and numerical studies of the rotation and waves active in B stars
\citep{Aerts2014,RogersMcElwaine2017}. Along with the fact that internal gravity
waves are also able to explain the angular momentum transport needed to bring
stellar models in agreement with the interior rotation rates derived from
asteroseismology \citep{Rogers2015,Aerts2019}, the search for and modelling of
non-radial pulsators born with a mass about 8\,M$_\odot$ was recently
intensified.

Thanks to high-precision uninterrupted space photometry of long duration, it has
become possible to derive a quantitative model-independent measurement of the
interior rotation of B stars from detected non-radial gravity modes and their
period spacing patterns
\citep{Triana2015,Papics2017,Kallinger2017,Szewczuk2018}.  Asteroseismic
modelling of such modes in B stars also offers estimation of the core
overshooting (and along with it the core mass) and of the level of macroscopic
chemical element transport in the envelope, as first established by
\citet{Moravveji2015,Moravveji2016}. Meanwhile, statistical methodology taking
into account the effects of the Coriolis force for the computation of the mode
frequencies and parameter degeneracies in the estimation was developed by
\citet{Aerts2018} for single stars and by \citet{Johnston2019} for
binaries. Those methodological frameworks require frequency precisions from
time-series data covering several times { the overall beating pattern due to
  the rotational and pulsational variability, as well as} unambiguous
identification of the degree and azimuthal order of the modes from frequency
splittings \citep{Triana2015} and/or from period spacing patterns
\citep{Papics2014,Papics2015,Papics2017, Szewczuk2018}.

Gravity-mode asteroseismology of B stars came on the radar after the first
discovery of period spacings expected for such modes in CoRoT space photometry
of the slowly rotating B3V star HD\,50230 \citep{Degroote2010}.  This gave the
first opportunity for asteroseismic estimation of the amount of core
overshooting based on gravity modes in a core-hydrogen burning star. Adopting
the description of convective penetration by \citet{Zahn1991},
\cite{Degroote2010} found a value between 0.2 and 0.3 local pressure scale
heights, for a star with $M\in [7,8]\,$M$_\odot$ and a central hydrogen mass
fraction of some 0.28, adopting solar metallicity. Moreover, the very small yet
clearly periodic deviations from a constant period spacing pattern implied a
transition region between the core overshoot zone and the radiative envelope,
where chemical mixing occurs at the level of
$D_{\rm mix}\simeq 10,000$\,cm$^2$\,s$^{-1}$. These first rough results based
on asteroseismic inference relying on gravity modes were confirmed by
\citet{Hendriks2019}, adopting a machine-learning approach for grid-based
asteroseismic modelling.

The 5-months long CoRoT monitoring turned out to be { barely} sufficient in
duration to discover gravity-mode period spacings and to establish inferences of
the mass, age, convective core overshooting and envelope mixing. { Those
  quantities can be assessed roughly, but only by 
fixing the initial hydrogen fraction and metallicity.} However, this
time base is insufficient to assess the interior physics properties with high
precision, unless one has additional independent information. This was the case
for the CoRoT target HD\,43317, which was found to be a magnetic gravity-mode
B-type pulsator rotating at half its critical rotation rate
\citep{Papics2012a,Briquet2013}.  In a combined polarimetric asteroseismic
study, \citet{Buysschaert2017} and \citet{Buysschaert2018} were able to derive a
high-precision estimate of the surface rotation frequency of this star, which
subsequently allowed forward asteroseismic modelling. Although limited in
capacity due to the limited number of identified modes, this revealed a
relatively young star of $\sim\!5.8\,$M$_\odot$ with a very weak core
overshooting of value $f_{\rm ov}\simeq\!0.004$, assuming an exponential
functional form as proposed by \citet{Freytag1996} and adopting the radiative
temperature gradient in the overshoot zone. It is tempting to assign the limited
overshooting to the inhibition of particle mixing from the magnetic field, but
the errors for $f_{\rm ov}\simeq\!0.004$ were too large to come to a firm
conclusion. The 5-month CoRoT light curve is not long enough to detect
sufficient independent gravity modes, due to the limited frequency resolution.

The {\it Kepler\/} space mission \citep{Gilliland2010} monitored stars during
four years, leading to a ten times better frequency resolution than achieved
with CoRoT. Thanks to these 4-year uninterrupted light curves, gravity-mode
asteroseismology of core-hydrogen burning stars is meanwhile an established
practise \citep{Aerts2018}. Nevertheless, applications to B stars are still
scarce because a limited sample was monitored and among those only a handful
were found to reveal appropriate and well identified modes so far
\citep{Aerts2019}.  In order to improve stellar evolution models of OB-type
stars from asteroseismology, the sample of modelled stars must be increased
drastically.  Studies of numerous {\it Kepler\/}, K2, BRITE, and TESS B-type
targets are ongoing or foreseen with this aim.  In anticipation of those,
and with the knowledge assembled from {\it Kepler\/} B-star asteroseismology, we
{ consider a CoRoT B-type target} in an attempt to detect and
interpret gravity-mode oscillations in this B2V star.

\section{The target star HD\,170580}

After a preparatory spectroscopic study by \citet{MorelAerts2007}, the B2
primary star of the multiple system HD\,170580 ($V$ magnitude 6.7) was accepted
as primary asteroseismology target for the sixth long run of the CoRoT space
mission \citep{Auvergne2009}.  { As such it was also included in the ESO Large
Programs for ground-based follow-up spectroscopy of CoRoT asteroseismology
targets with the HARPS spectrograph \citep{Poretti2013}, the data products of
which have been made available in the
SIMBAD archive \citep{Rainer2016}.} The choice
for this star, despite its poorly known multiplicity, was motivated by its
spectroscopic similarity to HD\,50230, in addition to it being situated in an
acceptable area on the sky for exoplanet hunting in the adjacent CCDs. Indeed,
just as HD\,50230, HD\,170580 was known to have a very low projected rotation
velocity, with $v\sin\!i=11\pm 1\,$km\,s$^{-1}$ derived by \citet{Lefever2010}
and $v\sin\!i=3\pm 1\,$km\,s$^{-1}$ by \citet{Bragan2012}, while
\citet{MorelAerts2007} found the overall rotational, pulsational, and
microturbulent spectral line broadening to be below 10\,km\,s$^{-1}$. Hence, the
prospects of finding another case with gravity-mode period spacings were good.
Its spectroscopic stellar parameters were found to be typical for its spectral
type: $T_{\rm eff}=20\,000\pm 1\,000$\,K, $\log g=4.10\pm 0.15$, where we note
that systematic uncertainties of $\log g$ occur and make this observable far
less reliable than $T_{\rm eff}$. Further, \citet{MorelAerts2007} found
$Z=0.0095\pm 0.0030$ and the NLTE abundance determinations by both
\citet{MorelAerts2007} and \citet{Lefever2010} revealed a He-weak star with
He/H=0.048$\pm$0.021 (by number).  This low helium fraction, along with the low
metallicity and very slow rotation, implies that atomic diffusion is active in
the star \citep[e.g.,][for a monograph]{Michaud2015}. As discussed by, e.g.,
\citet{Deal2016} and \citet{Aerts2018}, radiative levitation has not been
included in asteroseismic modelling of coherent non-radial oscillation modes in
OBA-type stars, because it is computationally too
demanding. \citet[][Fig.\,10]{Aerts2018} showed that ignoring atomic diffusion
typically leads to { high-order} gravity-mode frequency uncertainties near
0.001\,d$^{-1}$ for the mass range relevant for HD\,170580.  { However, the
  frequencies of low-order pressure modes are affected appreciably by atomic
  diffusion} (up to 1\,d$^{-1}$). We keep this in mind in the modelling
discussed below.

With its temperature and gravity, HD\,170580 is situated at the hot blue end of
the instability strip of the Slowly Pulsating B Stars (SPBs), in the overlapping
region with the $\beta\,$Cep strip \citep{Szewczuk2017}, so the star has the
potential to be a hybrid pressure and gravity mode pulsator.  CoRoT observations
of HD\,170580 were assembled during the third long run in the direction of the
Galactic centre (LRc03), and lasted some 170 days. Initial analyses of the CoRoT
and spectroscopic data revealed a different level and kind of variability than
for HD\,50230. The multiperiodicity of HD\,170580 occurs at a level of 100\,ppm
instead of 1\,000\,ppm for HD\,50230 and no obvious period spacing pattern could
be found, leaving the detected mode frequencies unidentified and hence forward
modelling impossible. With the knowledge gained from {\it Kepler\/} SPB
asteroseismology and with a Gaia DR2 parallax available, we now re-examine the
newly calibrated CoRoT light curve of HD\,170580 (Sect.\,2). Subsequently, we
analyse the high-resolution time-resolved spectroscopy (Sect.\,3), compare the
observational data with stellar models (Sect.\,4) and come to conclusions on the
overall variability behaviour of the star (Sect.\,5).

\section{CoRoT Photometry}

The CoRoT data of HD\,170580 { were} obtained during a long run, from BJD
2455293.39 until 24455463.84 (with an interruption of about 2 days between BJD
2455382.74 and BJD 2455384.57). The light curve spans a total of 170.5 days,
leading to a { Rayleigh limit of $1/170.5=0.00587$\,d$^{-1}$ and hence 
a resolving
  power for frequency analysis from a prewhitening procedure of
  $\sim 1.5/170.5=0.0088$\,d$^{-1}$ \citep{LoumosDeeming1978}.}  The CoRoT
asteroseismology CCDs operated with a time sampling of 32\,s. After removal of
bad points and outliers, the light curve contained $\sim 400\,000$ data points,
corresponding to an effective duty cycle of 88\%.

We first computed the Lomb-Scargle periodogram \citep{Scargle1982} of this
original light curve and found all variability to occur on time scales of hours
to weeks. Hence, the data were largely oversampled.  In order to reduce the
computation time of the frequency analysis, { we initially worked with two
  data sets: one with the original time steps and one for which the fluxes were
  averaged over each set of 10 consecutive data points. Since these two data
  sets led to the same results, we used the latter one for the detailed
  frequency analysis.}

\subsection{Detrending}

\begin{figure*}
\begin{center}
\rotatebox{270}{\resizebox{13cm}{!}{\includegraphics{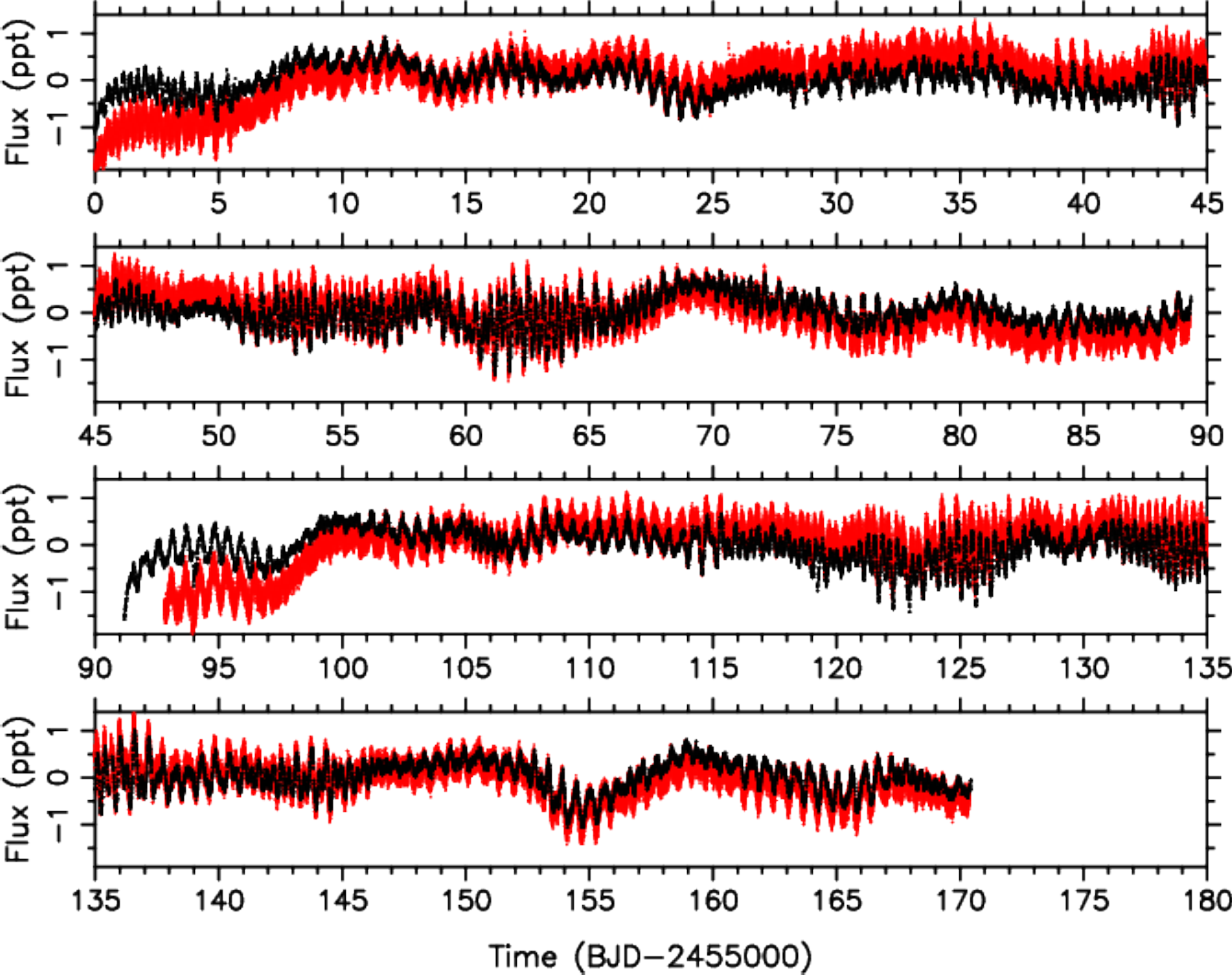}}}
\end{center}
\caption{{ Two versions of the CoRoT light curve of HD\,170580 (red: original
    light curve with 32\,s sampling, black: reprocessed light curve with 320\,s
    sampling and detrended by a 3rd-order polynomial). }}
\label{reduceddatafig}
\end{figure*}

Long-term trends in the data were removed by dividing the measured fluxes {
  by a polynomial fit to the light curve.  We used polynomials of various
  degrees and detrended in two ways for each of them: once for the light curve
  as a whole and once for each of the two separated sub-runs. In the end, we
  detrended by one global 3rd-order polynomial, as this delivered the best
  compromise in view of two properties:
\begin{enumerate}
\item an instrumental effect occurring the first days after the start
  and interruption of the run, showing a marked increase in the flux level;
\item
the intrinsic stellar variability, which occurs on time scales of both hours and
weeks. 
\end{enumerate}
After detrending, all fluxes were converted to ppm. Two versions of the
light curve are shown in Fig.\,\ref{reduceddatafig}. The red dots
represent the original light curve with 32\,s sampling, where  
part of the curve after the interruption was
omitted, while the black dots are the result after averaging to a ten times less
dense sampling and dividing by one global 3rd-order polynomial for the entire
light curve. }

All the different approaches for the detrending we tried in the end resulted in
the same values for the extracted frequencies of the star (within the errors).
The choice of the detrending only changed frequencies in the low-frequency
regime, where instrumental effects and stellar oscillations cannot be reliably
distinguished from each other. We chose to proceed with global detrending of the
overall light curve by { the best fit third-order polynomial derived from the
red curve in Fig.\,\ref{reduceddatafig} and work with the black curve for
  the frequency analysis. This method led to the highest} precision for the
derived frequencies and there was no good reason that justified separate
detrending for different parts of the light curve.

{ The Lomb-Scargle periodograms of both versions of the light curve shown in
  Fig.\,\ref{reduceddatafig} are plotted in the upper panel of
  Fig.\,\ref{freqspectrum} with the same colour convention. It can be seen that
  the detrending affects the lowest frequency regime. These
  frequencies must therefore be treated with caution.  In the rest of the paper,
  we rely on the version of the light curve shown in black in
  Fig.\,\ref{reduceddatafig}.}

\subsection{Frequency analysis}

\begin{figure}
\begin{center}
\rotatebox{270}{\resizebox{5.75cm}{!}{\includegraphics{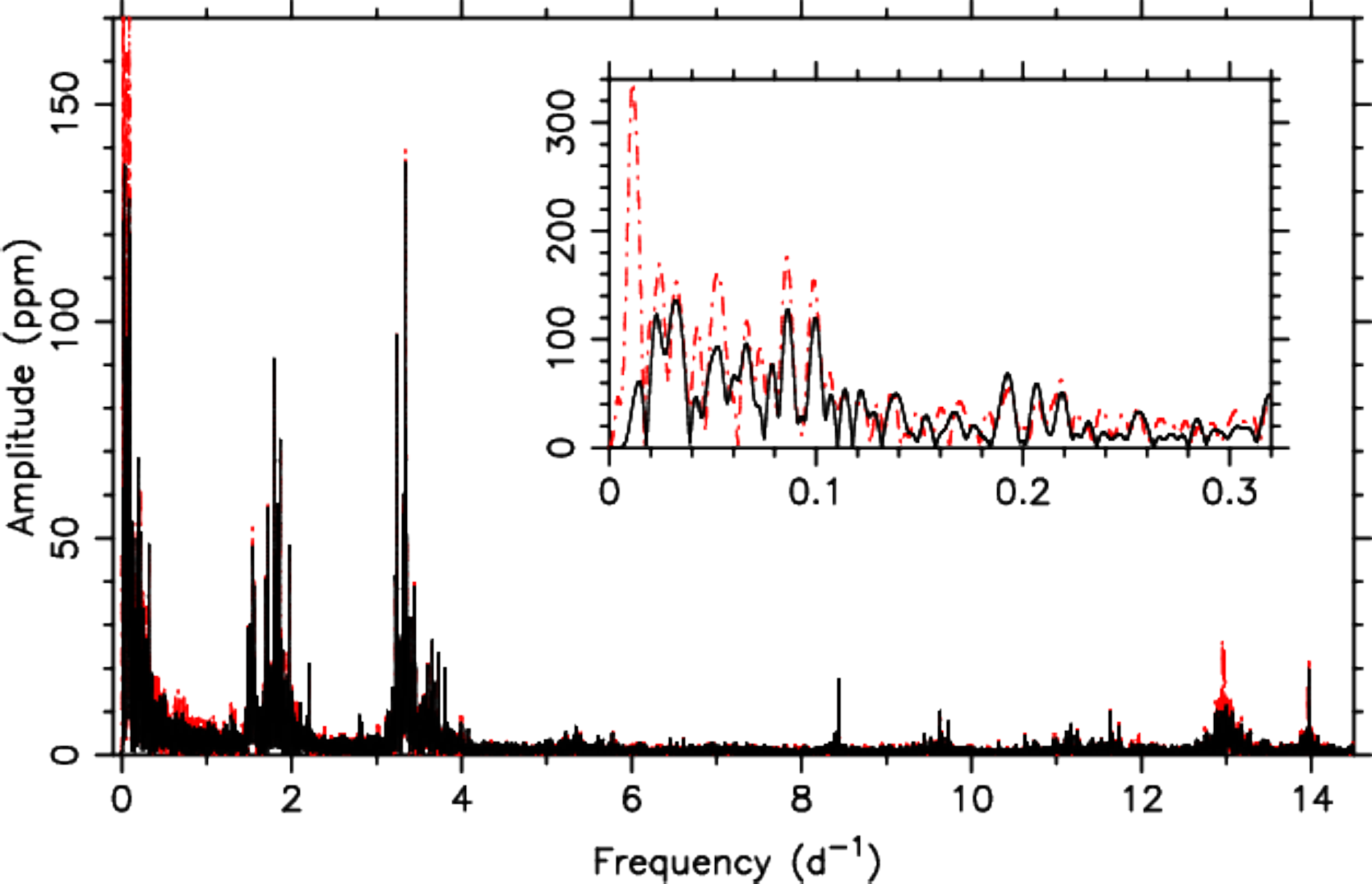}}}
\rotatebox{270}{\resizebox{5.75cm}{!}{\includegraphics{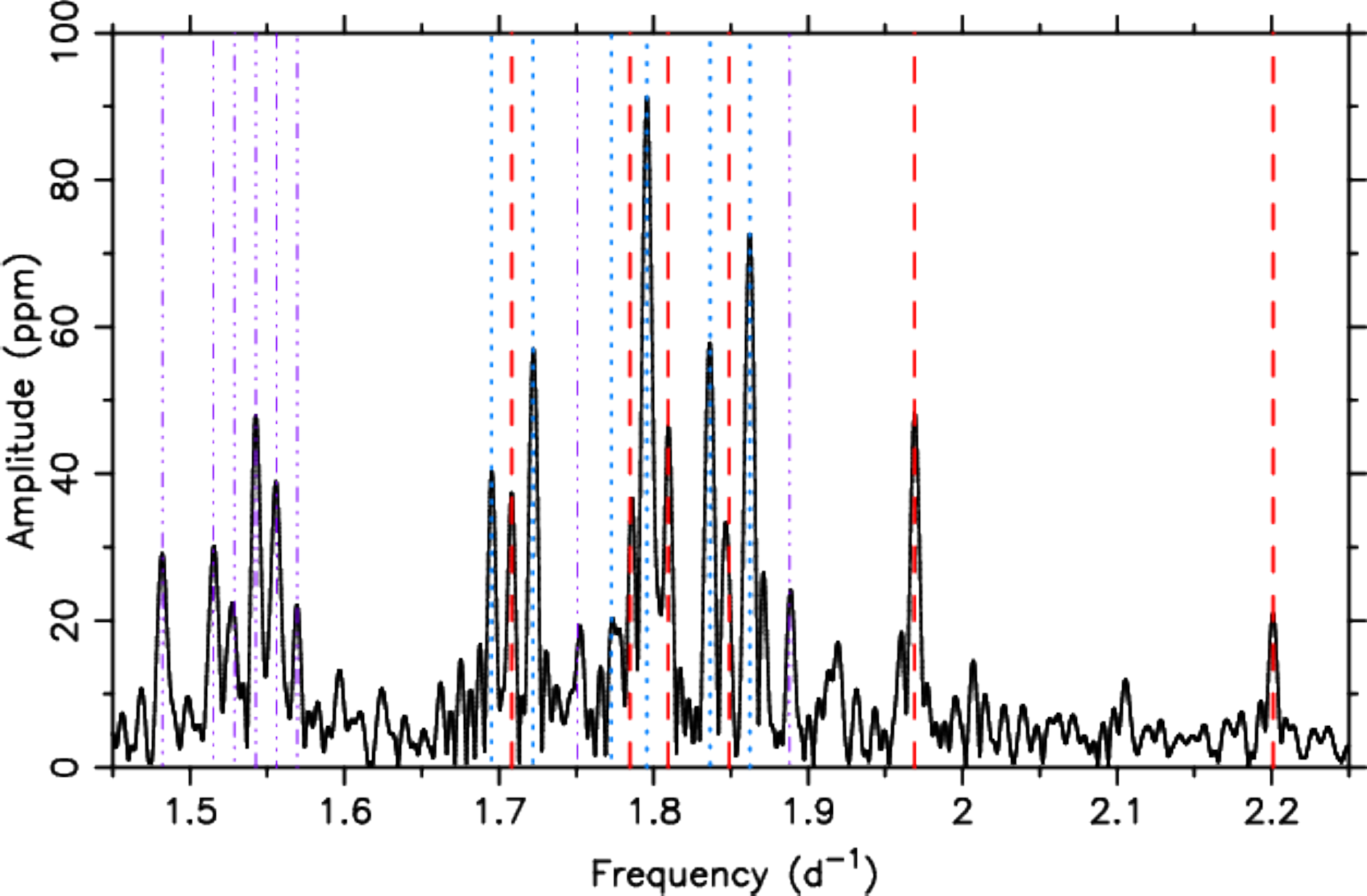}}}
\rotatebox{270}{\resizebox{5.75cm}{!}{\includegraphics{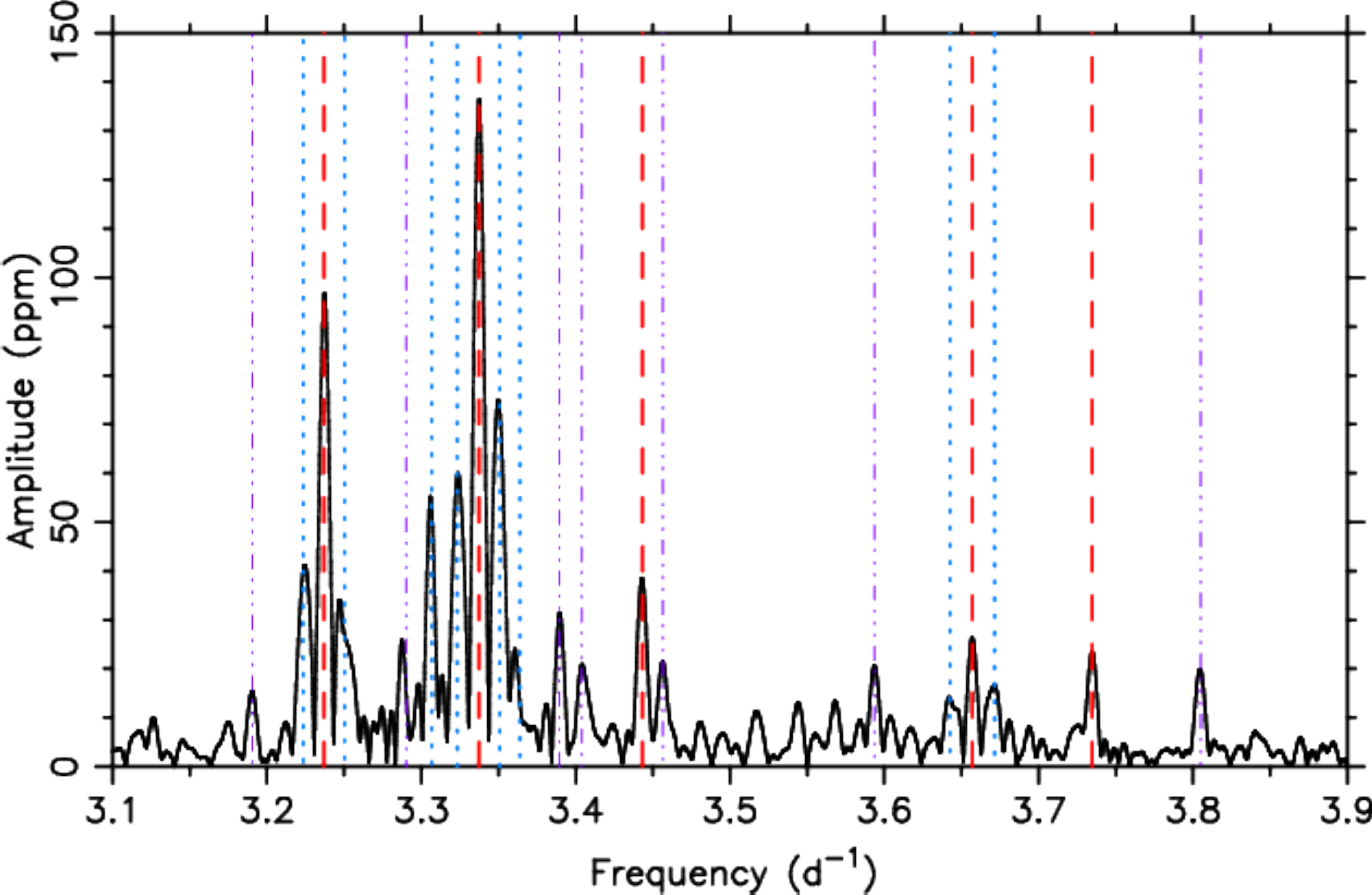}}}
\end{center}
\caption{ Upper panel: Lomb-Scargle periodograms of the CoRoT data shown in
  Fig.\,\ref{reduceddatafig} with the same colour convention.  Middle and lower
  panel: zoomed regions of interest, where the zonal mode frequencies used for
  the forward modelling are indicated by (red) dashed lines, the other multiplet
  components by (blue) dotted lines and combination frequencies by (purple)
  dashed-dot-dot-dot lines.}
\label{freqspectrum}
\end{figure}

In order to extract frequencies from the data, an iterative prewhitening
procedure was { applied. 
We used the method developed by \citet{Degroote2009}
  and further improved by \citet{Papics2012a}.  The Lomb-Scargle periodograms
  for the two versions of the light curve are shown in the top panel of
  Fig.\,\ref{freqspectrum}.  At each step of the prewhitening procedure, the
  frequency with maximum amplitude was identified in the periodogram.  This
  frequency value, along with all frequencies, amplitudes, and phases found from
  previous prewhitening stages, were fed into a non-linear least squares (NLLS)
  fit to the detrended light curve and the residual light curve was computed.
  Because we are dealing with periodic signals of amplitudes below 150\,ppm, it
  is justified to subtract the NLLS fit from the detrended light curve in ppm to
  start the search for the next frequency.

  Following this method, the prewhitening procedure was stopped when the
  signal-to-noise ratio (S/N) of a frequency in the amplitude spectrum,
  calculated in an interval of width 2\,d$^{-1}$ centered on the detected
  frequency, dropped below the canonical value of four \citep{Breger1993}.  This
  resulted in 177 extracted frequencies. However, \citet{Baran2015} showed that
demanding an amplitude above 5\, S/N is a more appropriate criterion to prevent
  over-interpretation for densely sampled space photometry with a time base of
  only a few months, as is the case for CoRoT or short-cadence K2 data.  We
  confirm this result, as many of the 84 frequencies with amplitude between four
  and five times the local noise level turn out to be unresolved with respect to
  higher-amplitude frequencies.  We therefore only considered the 93 frequencies
  that meet the more conservative criterion of having an amplitude above 5\, S/N.}
\begin{figure}[b!]
\centering
\rotatebox{270}{\resizebox{5.75cm}{!}{\includegraphics{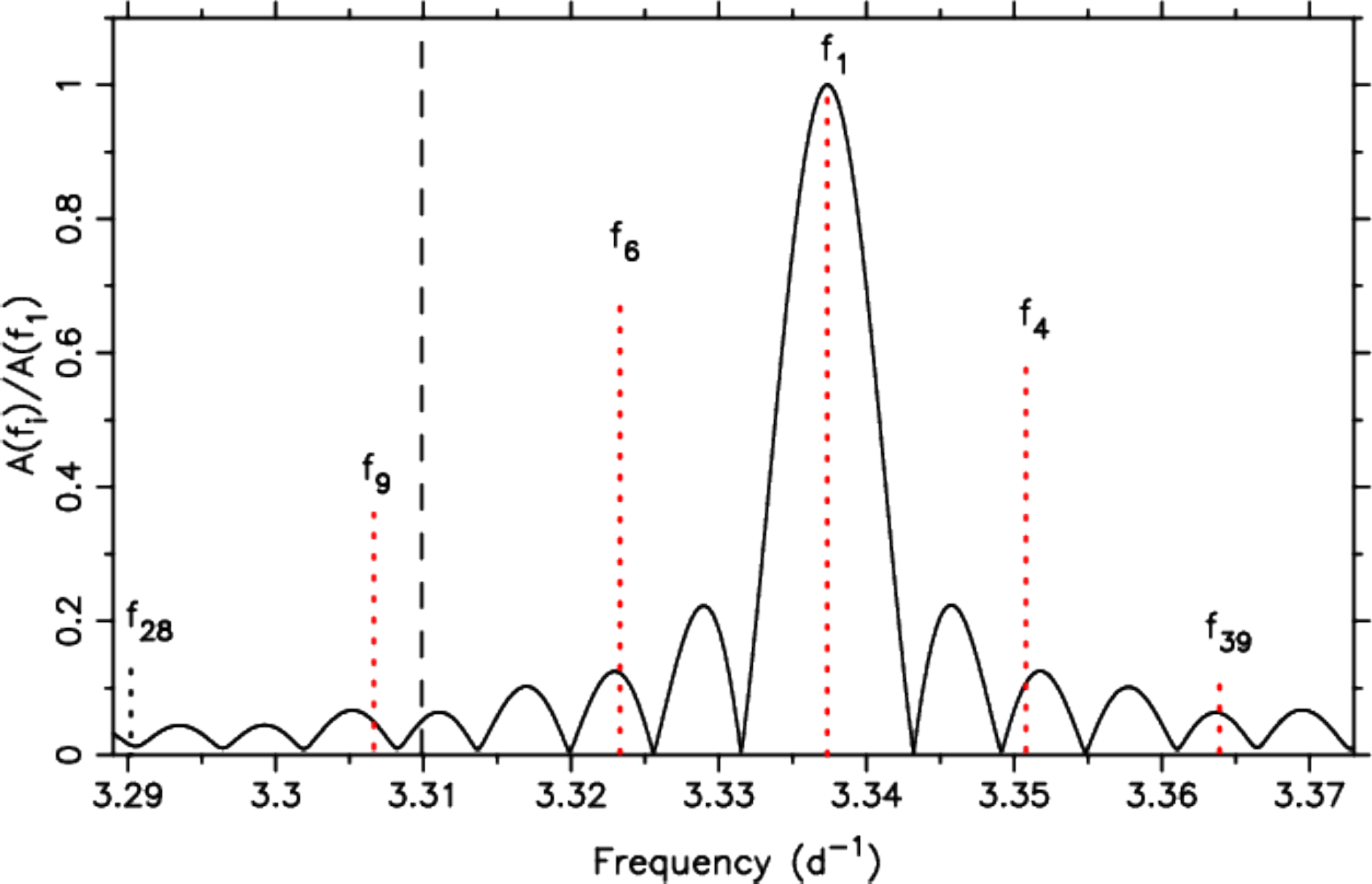}}}
\rotatebox{270}{\resizebox{5.75cm}{!}{\includegraphics{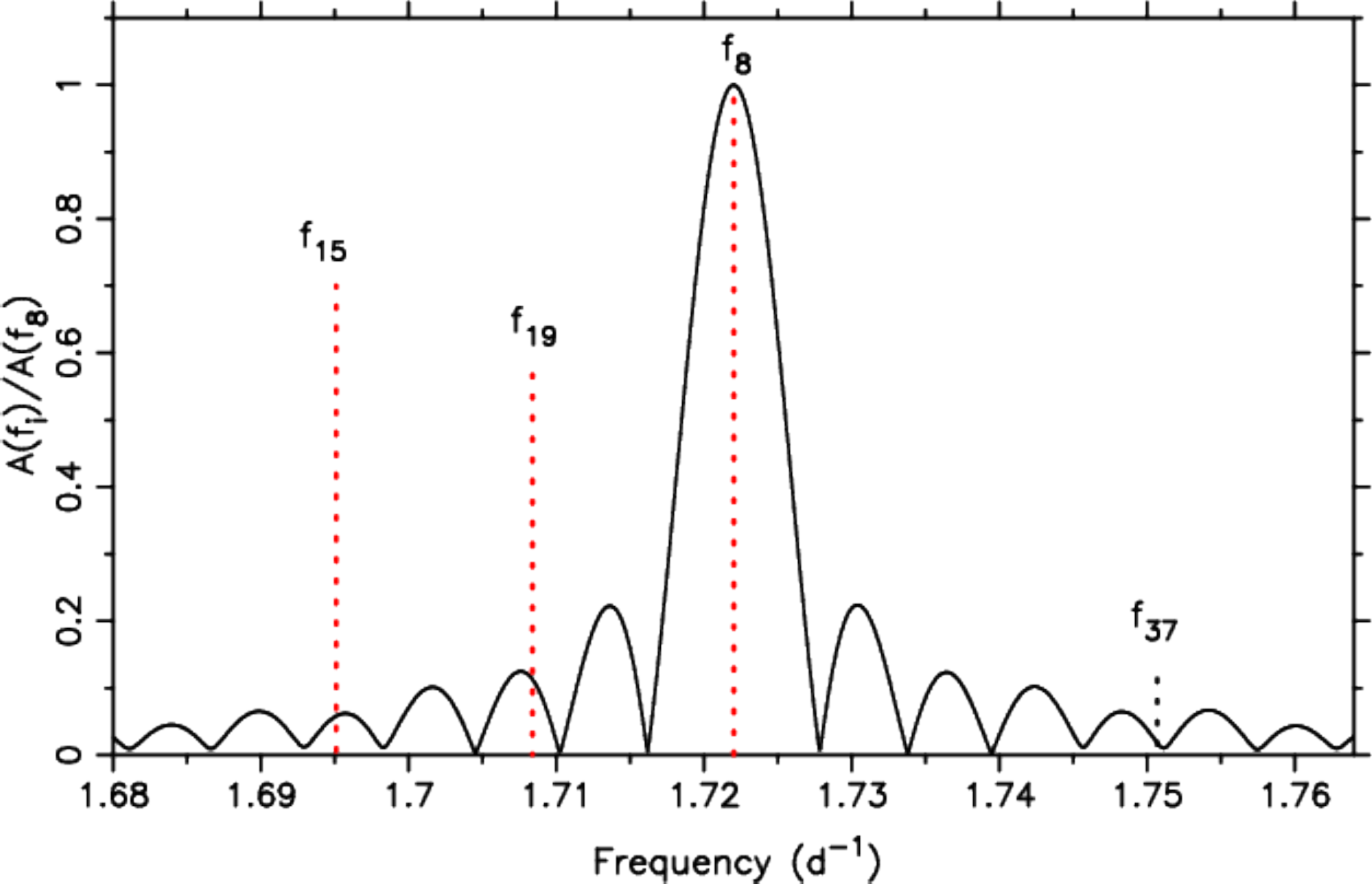}}}
\caption{Two zoomed parts of the amplitude spectrum, where the length of the
  $x$-axis equals ten times the Rayleigh limit. The spectral window (full line)
  is placed at the local dominant frequency (cf.\ Table\,\ref{freqtable}).
  Dotted lines indicate the extracted frequencies whose amplitude was normalised
  to the dominant one in that frequency regime (top: $f_1$, bottom:
  $f_8$). Black dotted lines indicate combination frequencies.  The dashed line
  in the upper panel is placed at the position where the fifth quintuplet
  component would occur in the case of equal splitting.}
\label{windows}
\end{figure}

{ Not all of the 93 frequencies that meet the adopted S/N criterion are
  intrinsic to the star.  Among them we found} $f_{W}=13.9693$\,d$^{-1}$, which
is a well-known frequency connected with the CoRoT satellite orbit
\citep{Auvergne2009}. { This frequency and its harmonics were} excluded for
asteroseismic interpretation.  { Further, the frequency
  $f_{41}=2.0069$\,d$^{-1}$ was found.  This frequency 
is due to the crossing of the South Atlantic
  Anomaly, which happens twice per day \citep{Poretti2010,Poretti2011}.}  We
listed it in Table\,\ref{freqtable} but we did not use it for further
interpretation.  { In addition,} all frequencies below 0.5\,d$^{-1}$ were
discarded. There may be oscillation frequencies { and/or harmonics of the
  rotation frequency in this region but they cannot be distinguished from
  frequencies induced by instrumental effects (cf.\ Fig.\,\ref{reduceddatafig}
  and the inset in the top panel of Fig.\,\ref{freqspectrum}).}  All frequencies
equal to a combination of frequencies with at least one component below
0.5\,d$^{-1}$ were also excluded, as well as those that differ less than the
Rayleigh limit from a previous one with larger amplitude.  Thanks to our
conservative approach in terms of S/N, we found only two frequencies that
survive this selection while not being fully resolved { according to the
  \citet{LoumosDeeming1978} criterion ($f_{16}-f_{17}\simeq0.0087$, cf.\
  Table\,\ref{freqtable}).}

After these strict selection criteria, the { 42} frequencies in addition to
$f_{41}$ listed in Table\,\ref{freqtable} are attributed to the star and further
explored. { \citet{Degroote2009} presented an extended analysis of the noise
  properties of the oversampled CoRoT asteroseismology data and concluded that
  the formal errors for the parameters resulting from a NLLS fit need to be
  corrected for the correlated nature of the data. As in that paper, we applied
  the procedure described in \citet{SchwarzenbergCzerny2003} to compute an
  average correction factor for the NLLS errors of the frequencies and
  amplitudes based on the residual light curve and obtained 1.84. This factor
  has been taken into account in the errors reported in Table\,\ref{freqtable}.
}

The 42 frequencies of the star listed in Table\,\ref{freqtable} occur in two
groups: one between 1.45\,d$^{-1}$ and 2.25\,d$^{-1}$ and another one between
3.1\,d$^{-1}$ and 3.9\,d$^{-1}$.  { Zoomed versions of the original Lomb-Scargle
  periodogram are shown in the middle and bottom panels of
  Fig.\,\ref{freqspectrum} for these two regions.}  There is only one extracted
frequency that does not belong to one of those two groups, namely
$f_{38} = 8.4402$\,d$^{-1}$.

Before interpreting the frequencies asteroseismically, it is necessary
to check for combination frequencies. Indeed, deviations of the star from
spherical symmetry, non-linear effects and interactions between different
oscillation modes can result in frequencies that are linear combinations of
their parent frequencies \citep{Degroote2009,Papics2012b,Kurtz2015}. Combination
frequencies do not necessarily correspond to eigenmodes of the star, as they
may also just be a consequence of the non-linearity of the light curve. For
asteroseismic modelling purposes, it is important to identify combination
frequencies and unravel if they can be due to an eigenfrequency of the star or not.
The CoRoT light curve of HD\,170580 does not reveal harmonics of single
frequencies (i.e.\ combinations of the form $c_{1}f_{i}$ with
$c_{1} \in \mathbb{N}$). The search for combinations of two frequencies (i.e.\
of the form $c_{1}f_{i} + c_{2}f_{j}$ with $c_{1}$, $c_{2} \in \mathbb{Z}$) was
split into two steps. For positive combinations ($c_{1}$, $c_{2} > 0$), we
allowed for a large number of values of $c_{1}$ and $c_{2}$. The result was 12
possible combinations of second order and one of third order. In the second
search, $c_{1}$ and $c_{2}$ were allowed to become negative. The frequency
difference between the two groups is such that a coincidental match becomes
increasingly likely for higher-order combinations, so we only searched for
differences between two frequencies. That way, 9 more second-order combinations
were found. All identified combination frequencies are listed as such in the
column ``Notes'' of Table\,\ref{freqtable}, where the less probable
identifications are flagged with a question mark.

The interpretation of the combination frequencies is not obvious, as the
frequency spectrum is dense in the two groups of frequencies.  Hence, the
probability of a coincidental mathematical match is correspondingly high,
keeping in mind { the Rayleigh limit}.
Given that harmonics do not occur and there is no
clear pattern for most combination frequencies, it is likely that most are just
coincidental mathematical matches. There are only two exceptions to that. First,
all frequencies of the first group below 1.6\,d$^{-1}$ are flagged as a
difference between a frequency of the second and a frequency of the first group
{ (cf.\ middle panel of Fig.\,\ref{freqspectrum})}.  As at least one
high-amplitude peak is involved in each of those combinations, this is probably
due to a physical effect connected with non-linear mode behaviour.  Second, the
only frequency not belonging to one of the two groups is a third-order
combination frequency ($f_{38}\simeq f_{3}+2f_{6}$).  We conclude from our
frequency analysis that there are two groups of independent oscillation mode
frequencies of the star, because none of the two groups can be fully explained
as combination frequencies of the other.

The frequency values of the independent modes of HD\,170580
(Table\,\ref{freqtable}) are lower than the typical frequency range for
low-order non-radial pressure modes in $\beta\,$Cep stars, while they are higher
than the frequencies of { high-order} gravity modes of SPBs, notably those
detected in the CoRoT data of HD\,50230 \citep{Degroote2012}. Shifts towards
higher frequencies for high-order gravity modes in SPBs can be caused by fast
rotation \citep{Papics2017,Buysschaert2018}, but given the low $v\sin\!i$ this
can only be invoked for HD\,170580 if we see it almost exactly pole on.
However, in that case one would not expect to detect the three components of
triplets while we do so (cf.\ Fig.\,\ref{freqspectrum}). Moreover, a fast
rotating star is not expected to be subjected to the effects of atomic
diffusion, while { this phenomenon is active in HD\,170580 given its low
  measured surface helium and metal abundances.} These facts make it more
probable that the frequencies of the star are due to low-order modes seen from a
viewing angle that does not coincide with its pole.  In that case the zonal mode
frequency values imply a more massive and more evolved star compared with
HD\,50230.  \citet{Hendriks2019} found $M\sim 7.9\,$M$_\odot$ and a central
hydrogen mass fraction of $X_c\sim 0.17$ for HD\,50230. We keep this information
in mind in the mode identification and modelling described below.

\begin{table}
  \caption{ Extracted frequencies of HD\,170580 with amplitude above 
5 times the local
    noise level. The formal errors from a NLLS fit were corrected for the
    correlated nature of the data following \citet{SchwarzenbergCzerny2003} and
    are 
    indicated in brackets at the level of the last digits. The significance  (in terms
    of S/N) was computed over an interval of 2\,d$^{-1}$ centered on the
    prewhitened frequency. The frequencies in italic are omitted in the
    interpretation -- see text for discussion of the notes. The last row lists the 
    dominant frequency found in the HERMES spectroscopy, while $f_3$ was
    recovered in the HARPS spectra.}
\tabcolsep=3pt
\begin{tabular}{llrrc}
\hline \hline
\# & f (d$^{-1}$) & A (ppm) & S/N & Notes \\ \hline
$f_1$ & 3.3373(4) & 148(17) & 23.8 & quintuplet\\
$f_2$ & 3.2371(5) & 103(16) & 11.4 &  triplet \\
$f_3$ & 1.7957(6) & 82(15) & 11.4 & triplet, in HARPS spectra\\
$f_4$ & 3.3509(6) & 87(15) & 11.4 & quintuplet \\
$f_5$ & 1.8622(6) & 72(14) & 11.4 & triplet \\
$f_6$ & 3.3233(7) & 71(15) & 11.4 & quintuplet\\
$f_7$ & 1.8364(7) & 61(13) & 11.9 & triplet \\
$f_8$ & 1.7221(8) & 53(13) & 12.2 & triplet \\
$f_9$ & 3.3066(9) & 54(14) & 11.6 & quintuplet \\
$f_{10}$ & 1.9691(9) & 49(14) & 12.4 & \\
$f_{11}$ & 1.5426(7) & 48(10) & 11.2 & $f_{1}-f_{3} $\\
$f_{12}$ & 3.2237(10) & 46(14) & 11.9 & triplet\\
$f_{13}$ & 1.5559(6) & 42(8) & 8.5 & $f_{4}-f_{3} $\\
$f_{14}$ & 3.4432(11) & 39(14) & 12.2 & \\
$f_{15}$ & 1.6950(7) & 36(8) & 8.6 & triplet \\
$f_{16}$ & 1.8095(6) & 46(8) & 8.8 & \\
$f_{17}$ & {\it 1.8008(6)} & {\it 41(8)} & {\it 8.9} & {\it unresolved}\\
$f_{18}$ & 3.2505(12) & 37(13) & 12.5 & triplet, $f_{13}+f_{15}$\\
$f_{19}$ & 1.7083(8) & 31(8) & 9.1 & triplet \\
$f_{20}$ & 1.5151(8) & 33(8) & 9.3 & $f_{2}-f_{8}$\\
$f_{21}$ & 1.4822(7) & 28(6) & 9.0 & $f_{6}-f_{7}$\\
$f_{22}$ & 3.3895(9) & 29(8) & 9.1 & $f_{7}+f_{13}$?\\
$f_{23}$ & 3.6568(9) & 29(8) & 9.1 & triplet, $f_{3}+f_{5}$\\
$f_{24}$ & 1.8495(10) & 24(8) & 9.4 & triplet\\
$f_{25}$ & 1.5285(8) & 24(6) & 7.4 & $f_{1}-f_{16}$\\
$f_{26}$ & 1.7849(9) & 22(6) & 7.2 & triplet \\
$f_{27}$ & 3.7348(11) & 23(8) & 9.3 & \\
$f_{28}$ & 3.2902(13) & 20(8) & 9.3 & $f_{16} + f_{21}$?\\
$f_{29}$ & 3.5936(13) & 20(8)& 9.3 & $f_{3}+f_{17}$\\
$f_{30}$ & 3.8053(12) & 22(8) & 9.3 & $f_{7}+f_{10}$\\
$f_{31}$ & 3.6715(13) & 20(8) & 9.3 & triplet, $f_{5}+f_{16}$? \\
$f_{32}$ & 2.2010(9) & 21(6) & 7.5 & \\
$f_{33}$ & 3.4567(12) & 21(8) & 9.3 & $f_{10}+f_{21}$? \\
$f_{34}$ & 3.4040(12) & 21(8) & 9.3 & $f_{15} + f_{19}$?, $f_{5}+f_{11}$? \\
$f_{35}$ & 1.5694(6) & 18(3) & 6.8 & $f_{4}-f_{19}$ \\
$f_{36}$ & 1.8880(8) & 23(6) & 7.5 & $f_{27}-f_{24}$? \\
$f_{37}$ & 1.7510(10) & 19(6) & 7.6 & $f_{14}-f_{15}$? $f_{2}-f_{21}$? \\
$f_{38}$ & 8.4401(28) & 19(16) & 11.3 & $\simeq f_{3} + 2 f_{6}$ \\
$f_{39}$ & 3.3641(11) & 18(6) & 7.4 & quintuplet \\
$f_{40}$ & 1.7727(7) & 16(3) & 6.8 & triplet \\
$f_{41}$ & {\it 2.0069(6)} & {\it 17(3)} & {\it 6.9} & {\it satellite}\\
$f_{42}$ & 3.1905(13) & 14(6) & 7.7 & $f_{19}+f_{21}$?\\
$f_{43}$ & 3.6433(12) & 15(6) & 7.7 & triplet, $f_{3}+f_{24}$?\\
\hline
$f_{S1}$ & 2.6700 & & & in HERMES spectra \\
&&&& alias of $f_{31}$\\
\hline
\end{tabular}
\label{freqtable}
\end{table}

Examination of the frequencies not due to combinations in Table\,\ref{freqtable}
did not reveal any obvious gravity-mode period spacing pattern as it has been
found for various other SPB pulsators from space photometry \citep[cf.\
Table\,B1 in][for an update and typical values]{Szewczuk2018}. This is not so
surprising because the detected frequencies of HD\,170580 likely correspond with
low-order pressure or gravity modes and these are not expected to reveal period
spacing patterns. However, we do find { six 
  multiplets. Five of these are triplets with splitting values ranging from
  $0.0108$ to $0.0147$\,d$^{-1}$: $(f_{15},f_{19},f_{8})$ with spacings of
  0.0133$\pm 0.0015$\,d$^{-1}$ and 0.0138$\pm 0.0016$\,d$^{-1}$,
  $(f_{12},f_{2},f_{18})$ with spacings of 0.0134$\pm 0.0015$\,d$^{-1}$ and
  0.0134$\pm 0.0017$\,d$^{-1}$, $(f_{7},f_{24},f_{5})$ with spacings of
  0.0131$\pm 0.0017$\,d$^{-1}$ and 0.0127$\pm 0.0016$\,d$^{-1}$,
  $(f_{40},f_{26},f_{3})$ with spacings of 0.0122$\pm 0.0016$\,d$^{-1}$ and
  0.0108$\pm 0.0024$\,d$^{-1}$, and $(f_{43},f_{23},f_{31})$ with spacings of
  0.0135$\pm 0.0021$\,d$^{-1}$ and 0.0147$\pm 0.0022$\,d$^{-1}$.  
The other one
  is a set of four components of a quintuplet $(f_{6},f_1,f_{4},f_{39})$, with
  spacings of 0.0140$\pm 0.0011$\,d$^{-1}$, 0.0135$\pm 0.0010$\,d$^{-1}$, and
  0.0131$\pm 0.0017$\,d$^{-1}$.}  These frequency splittings { have values
  between 1.9 and 2.5 times the Rayleigh limit and are indicated as blue dotted
  lines in Fig.\,\ref{freqspectrum}.} The detected frequencies for two of the
multiplets have been plotted with respect to the spectral window in
Fig.\,\ref{windows}, from which we see that the frequency $f_9$ is shifted by
0.0167$\pm 0.0016$\,d$^{-1}$ from $f_6$.  The difference with the splitting
value of the other components (cf.\ red dotted and black dashed line in
Fig.\,\ref{windows}) is only about half the Rayleigh limit so $f_9$ { cannot
  formally be distinguished from the leftmost quintuplet
  component.}  From
Fig.\,\ref{windows} we deduce that the dominant spectral window frequency peak
of the first sidelobe occurs with an amplitude of 22\% at 1.4 times the Rayleigh
limit.  In particular, $f_4$ and $f_6$ occur in the second sidelobe of the
spectral window centered at $f_1$, which has an amplitude less than 5\%, leaving
no doubt about the reality of the frequencies indicated as multiplets in
Table\,\ref{freqtable}. { However, as indicated in Table\,\ref{freqtable} several of
  the triplet components with low amplitude differ less than the Rayleigh limit
  from combination frequencies. We keep this in mind in the modelling discussed
  in Sect.\,6.}

We conclude to have found evidence of { six} multiplets in the CoRoT data of
HD\,170580.  These { six multiplets} are compatible with rotational splitting
of dipole or quadrupole low-order modes.  Interpreting them as such and adopting
the Ledoux approximation, for which the splitting values between adjacent
multiplet frequencies is given by $(1-C_{nl})\Omega$ with $\Omega$ the angular
rotation frequency under the assumption of rigid rotation \citep{Ledoux1951} ,
leads to { $\Omega\in [0.0108,0.0294]\,$d$^{-1}$ because low-order pressure
  and gravity modes have $C_{nl}\in [0.0,0.5]$
  \citep[e.g.,][Chapter\,3]{Aerts2010}.  This implies that HD\,170580 has a
  rotation period averaged throughout the star between roughly 34 and 93\,d.  We
  refine this interval below, after modelling of the zonal modes, keeping in
  mind that low-order pressure-mode splittings are model dependent (through
  their model dependency of $C_{nl}$), while high-order gravity-mode splittings
  are not depending on stellar models because their $C_{nl}\simeq 1/[l(l+1)]$
  \citep[see][for a more extended
  explanation]{Aerts2010,Aerts2019}. Irrespective of possible model dependence
  for the derivation of $\Omega$,} it concerns slow rotation compared to the
periods of its oscillation modes, with $f_i/\Omega\!<\!0.02$ for all
$i=1,\dots,42$ in Table\,\ref{freqtable} and implies that the Coriolis and
centrifugal forces can be ignored when fitting the detected oscillation
frequencies.

\section{Luminosity from Gaia DR2}

We developed a procedure to derive the luminosity of pulsating B stars observed
with space photometry from the Gaia DR2 data release \citep{Luri2018}.\footnote{
  This procedure will be described in detail in a separate paper Pedersen et
  al., in preparaton, and is omitted here.}  The luminosity of HD\,170580 was
determined using the distance { of the star provided by
  \citet{Bailer-Jones2018} as $d=408.83$\,pc, with a $1\sigma$ confidence
  interval of $[399.30,418.83]$\,pc, on the basis of Gaia DR2 parallax
  measurements} \citep{Brown2018,Luri2018}.  Based on the apparent magnitudes
measured in the Johnson UBV, Cousins RI, TYCHO BV and 2MASS JH filters, we
calculated the apparent bolometric magnitude $m_\text{bol} = 3.48\pm0.15$ after
correcting for extinction. The estimated extinction in each filter was obtained
using the corresponding effective wavelengths as input for the York Extinction
Solver \citep{McCall2004}, assuming the reddening law by
\citet{Fitzpatrick1999}, with $R_V = 3.07$ for Vega and a reddening of
$E(B-V) = 0.41\pm0.04$ obtained from the reddening map published in
\citet{Green2018}. In order to convert the apparent magnitudes into the
bolometric scale, synthetic bolometric corrections were calculated for each
filter from \citet{Castelli2004} model atmospheres.
These bolometric corrections were calibrated such as to achieve the absolute
filter magnitudes of the Sun, assuming $M_{\text{bol},\odot} = 4.74$ as in
\citet{Torres2010}.
\begin{figure}
\centering
\rotatebox{270}{\resizebox{8.5cm}{!}{\includegraphics{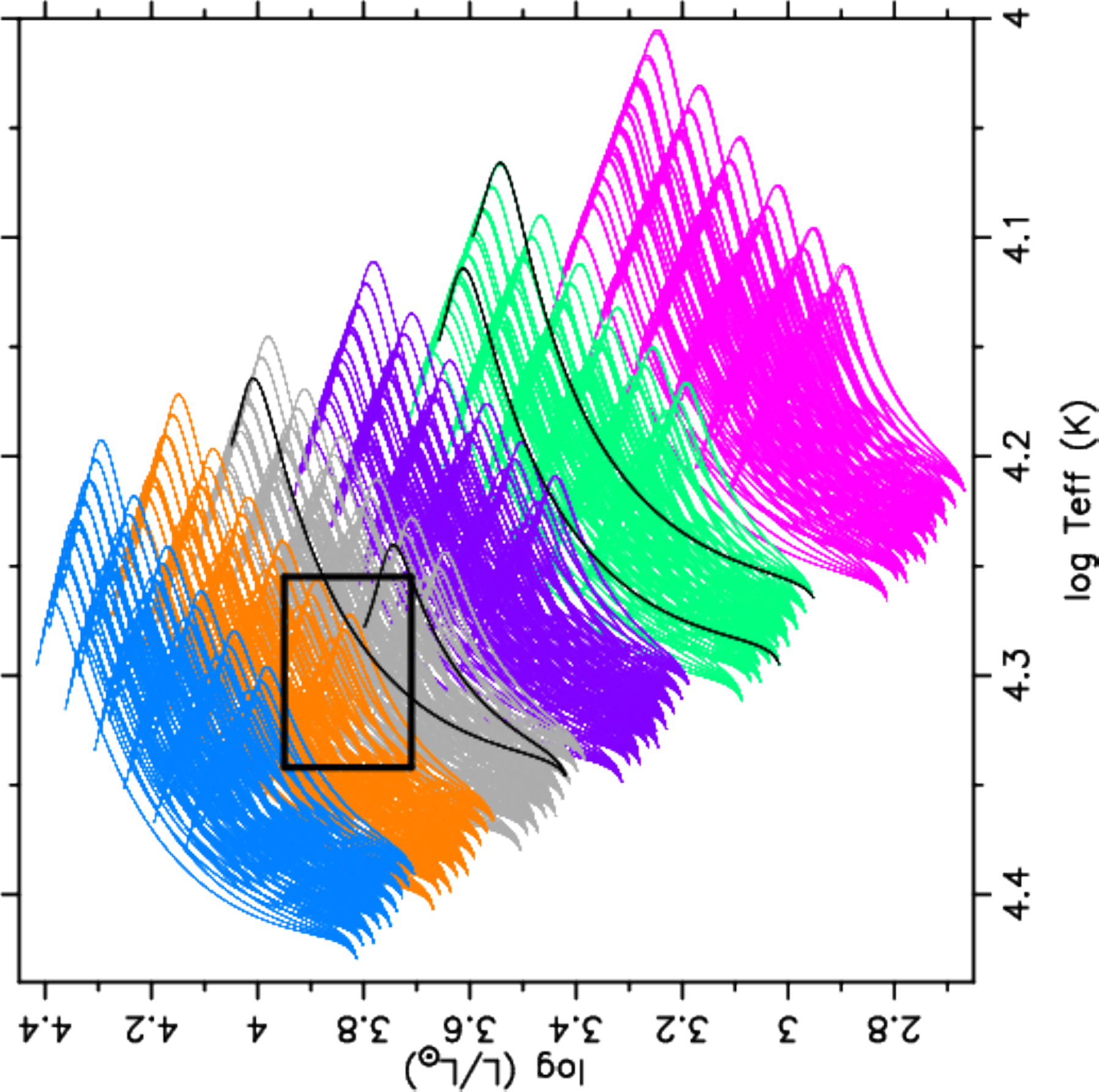}}}
\caption{Herzsprung-Russell Diagram with the 240 evolutionary tracks per mass in the
  SpaceInn grid for the masses 5 (pink), 6 (green), 7 (purple), 8 (grey), 9 (orange), 10
  (blue)\,M$_\odot$, along with the $2\sigma$ error box of HD\,170580 based on
  Gaia DR2 data and spectroscopy. The pulsation modes of the four tracks in
  black are shown in
  Fig.\,\ref{mode-diagram}.}
\label{HRD}
\end{figure}

In this way, we obtained $\log(L/L_\odot)=3.83\pm 0.06$, where we must
keep in mind
that the multiplicity was not yet treated in Gaia DR2 data and this may imply an
unknown systematic uncertainty for $L$.  Using this result along with
$T_{\rm eff}=20\,000\pm1\,000\,$K from \citet{MorelAerts2007}, we indicate the
position of HD\,170580 in the Hertzsprung-Russell (HRD) diagram using a
$2\sigma$ error box in Fig.\,\ref{HRD}, along with evolutionary tracks for
stellar models with masses { 5, 6,} 7, 8, 9, 10\,M$_\odot$ (these will be discussed
below).  Irrespective of the precise mass, we find that the star no longer
resides near the ZAMS but refinement of its evolutionary stage requires seismic
modelling. The Gaia luminosity and spectroscopic effective temperature leads to a
radius in the range $R\in[5,10]\,$R$_\odot$ (2$\sigma$-level). The
spectroscopic $\log\!g$ is of modest precision and therefore hardly implies a
useful restriction on the mass. We obtain $M\in[6,93]\,$M$_\odot$ at
2$\sigma$-level.

\section{Time-series spectroscopy}

{ In general, white-light space photometry alone may be insufficient for
  asteroseismic modelling based on low-order heat-driven oscillation modes,
  because these do not necessarily reveal patterns that lead to the
  identification of the mode wavenumbers $(l,m)$. For this reason, the CoRoT
  mission was accompanied by carefully planned time-series spectroscopy
  campaigns.}
We have collected a total of 208 spectra of HD\,170580, of which 60 were
obtained during 27.3\,d in June and July 2009 using the High Accuracy Radial
velocity Planet Searcher (HARPS) spectrograph at the ESO La Silla 3.6-m
telescope in Chile. This fibre-fed \'echelle spectrograph covers a wavelength
range of $378 - 691 \ \text{nm}$ and has a resolving power of $R = 115000$
\citep{Mayor2003}. The achieved S/N ratio ranges from 70 to 200. The remaining
148 spectra were assembled with the 1.2-m Mercator Telescope on La Palma, Spain,
using the High Efficiency and Resolution Mercator \'Echelle Spectrograph
(HERMES) in high-resolution mode ($R = 85000$) \citep{Raskin2011}. HERMES covers
the wavelength range $377-900 \ \text{nm}$. One spectrum was obtained in June
2009 and all others during 63.9\,d in May-July 2010, i.e., approximately one year
after the HARPS spectra. The total time base of the merged HARPS and HERMES
spectroscopy is 394\,d, giving a Rayleigh limit of 0.0025\,d$^{-1}$. 
All spectra were wavelength calibrated using Th-Ar
spectra obtained at the beginning and/or end of the night of observation, using
the HERMES and HARPS reduction pipelines. The spectra were normalised using the
method by \citet{Papics2012a}.

\subsection{Frequency analysis}

Scargle periodograms with the aim of extracting pulsation signals and performing
mode identification were computed using both the moment method
\citep[e.g.][]{Balona1986,Aerts1992,Briquet2003} and Pixel-by-Pixel method
\citep[e.g.][]{Telting1997,Zima2008} applied to individual spectral lines: the
\ion{Si}{iii} $4560 \AA$ triplet, the \ion{Si}{ii} $4130\AA$ doublet and the \ion{He}{i}
$6678 \AA$ line. These lines have previously been shown to be optimal for
frequency extraction and mode identification of $\beta$\,Cep stars
\citep{Aerts2003}, SPBs \citep{Aerts1999} and B-type fast rotating stars
\citep[e.g.][]{Balona1999}, respectively. In practice, we carried out the
analyses of the spectral line variations using the software package
\textsc{famias} developed by \citet{Zima2008}.

Considering each of the above mentioned three spectral lines separately, the
first step of the analysis included extracting the spectral line from the
normalised spectra including a small part of the continuum on either side of the
lines for S/N estimate. The wavelengths were converted to velocities using the
rest wavelength of the spectral line as the zero-point. After this conversion,
we calculated the moments of the lines and their Fourier spectra. In this way,
we did not find any significant variability. This is not so surprising, given
that line-profile variability was also not detected for HD\,50230
\citep{Degroote2012}, although the dominant modes of this star have amplitudes
ten times higher than those of HD\,170580.

A second attempt at extracting pulsation signals from the spectra was made by
computing least-squares deconvolution (LSD) profiles following the method
originally developed by \citet{Donati1997}. Such LSD profiles are a great asset
to detect low-amplitude variability, even if it cannot be found in individual
spectral lines. This was shown to be paticularly effective for g-modes in
$\gamma\,$Dor stars \citep[e.g.,][]{DeCat2006}.  We applied the more recent LSD
formalism of \citet{Tkachenko2013} to the HARPS and HERMES spectra separately,
including all spectral lines except the hydrogen and helium lines. The LSD
profiles are obtained by cross-correlating the lines in the observed spectra
with a set of delta functions constructed from an atomic line list, containing
information on rest wavelengths and theoretical line strengths. The line list
was computed using $T_\text{eff} = 20\,000 \ \text{K}$, $\log g = 4.1$ and solar
metallicity following the results by \citet{MorelAerts2007}.  For the further analysis, we
excluded one bad HARPS LSD profile and four bad HERMES LSD profiles.  Visual
inspection of the residual HARPS and HERMES LSD profiles reveales low-amplitude
line-profile variability.
\begin{figure}
\centering
\includegraphics[width=\linewidth]{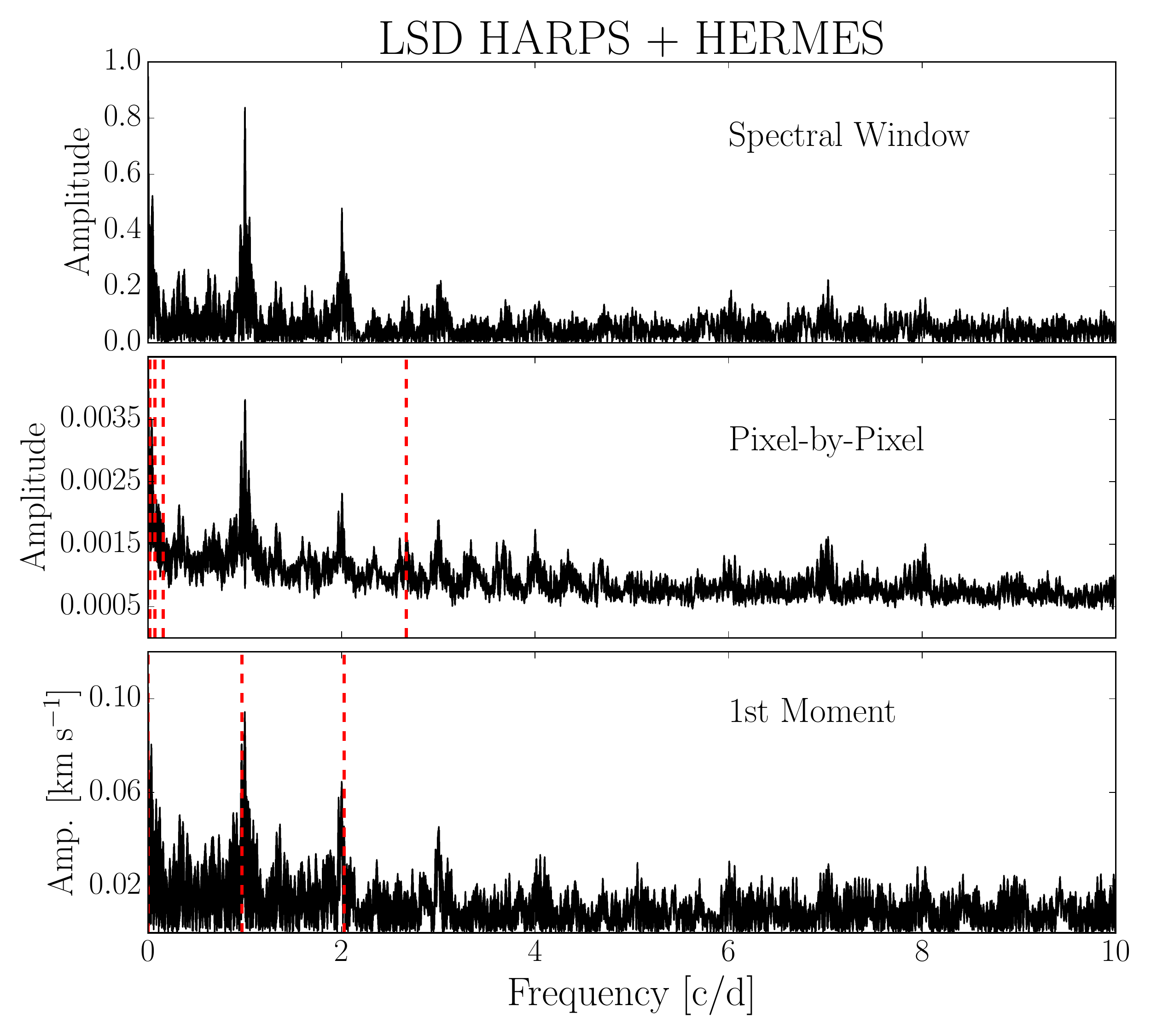}
\caption{Fourier spectra based on the Pixel-by-Pixel method (second panel) and
  the 1st moment (third panel) for the combined HARPS and HERMES LSD
  profiles. The top panel shows the spectral window.  Vertical dashed lines
  indicate the positions of frequencies with $S/N \geq 4$ during prewhitening.}
\label{FourierLSD}
\end{figure}

\begin{figure*}
\centering
\includegraphics[width=0.45\linewidth]{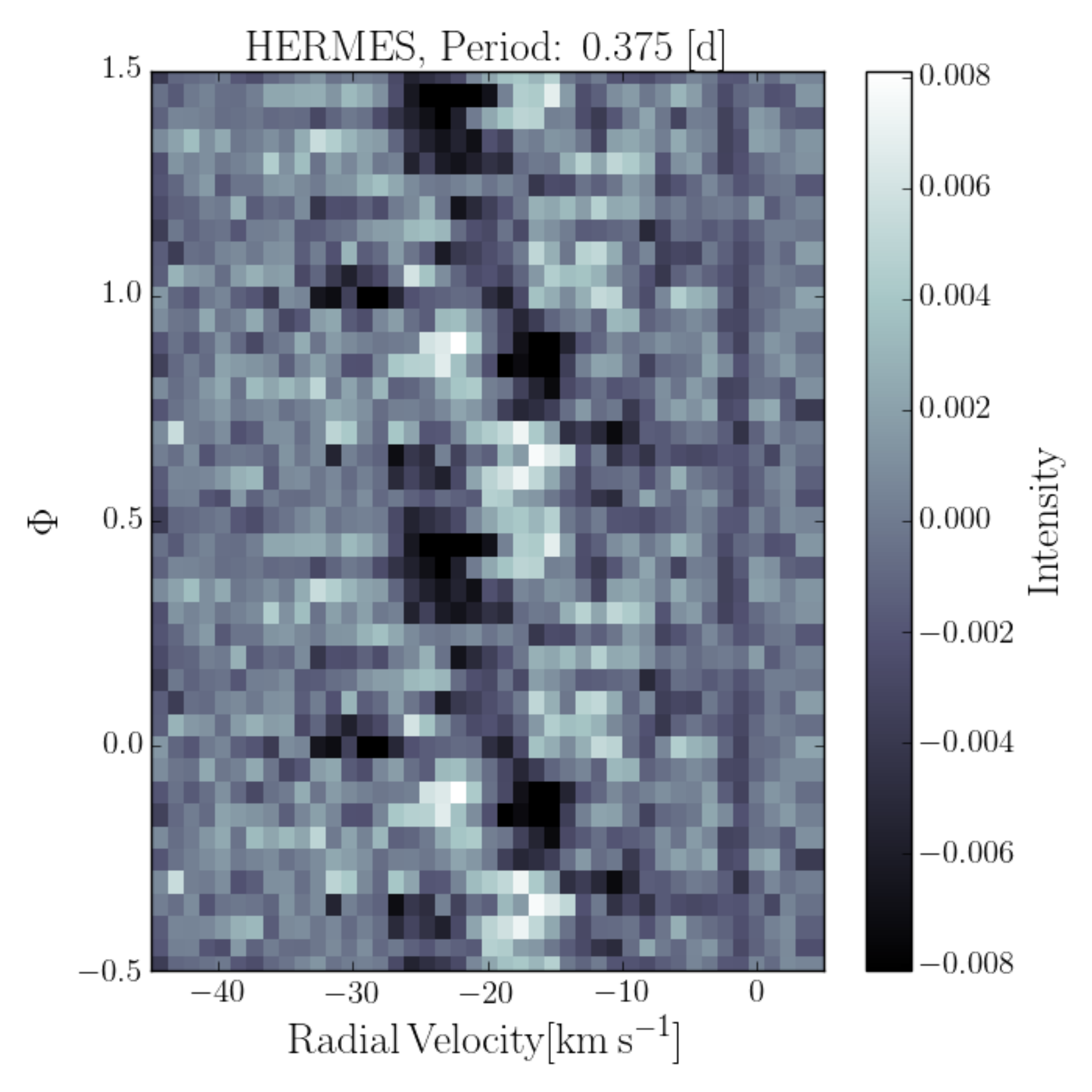}
\includegraphics[width=0.45\linewidth]{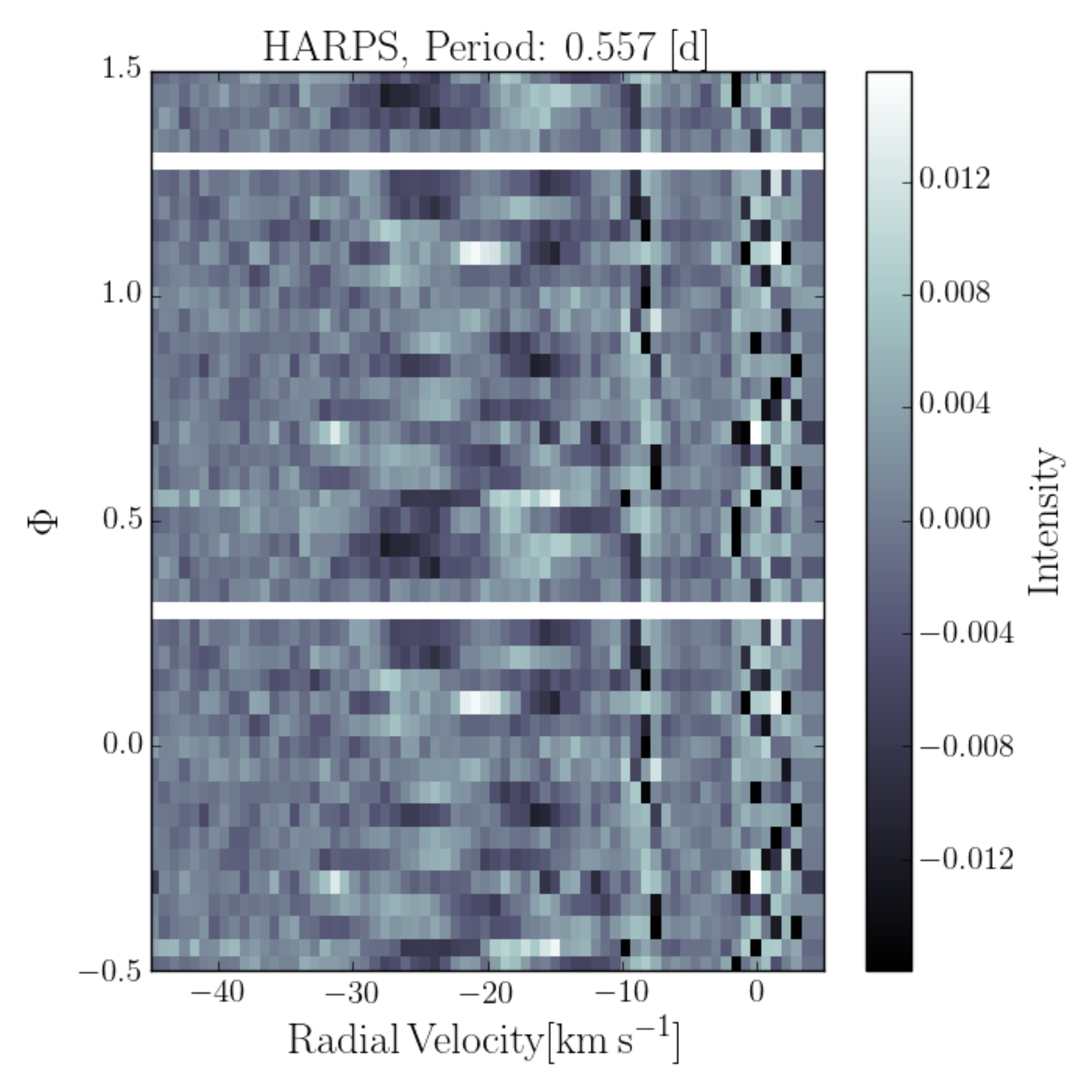}
\caption{Phase folded residual LSD profiles for the HERMES (left, for
  $f_{S1} = 2.6700 \ \text{d}^{-1}$) and HARPS (right,
  $f_3 = 1.79567 \ \text{d}^{-1}$) spectra. The phase folded residual profiles
  are averaged over an 0.05 phase interval and plotted over twice the phase
  interval for better visibility. The
  straight lines near -10 and 0\,km,s$^{-1}$ are caused by telluric lines.}
\label{LSDphasefold}
\end{figure*}

Using the LSD profiles of both the HARPS and HERMES spectra we recomputed
Fourier spectra based on the velocity moments and Pixel-by-Pixel method as
described above. The results are displayed in Fig.\,\ref{FourierLSD} for the
Pixel-by-Pixel method and the 1st moments along with the spectral window of the
data. By prewhitening the data according to the highest amplitude peaks in the
Fourier spectra, we detected five candidate frequencies from the Pixel-by-Pixel
computed Fourier spectra and three candidates for the 1st moments.  However,
most of these are connected with the time sampling and cannot be accepted as
intrinsic to the star, except for $f_{S1}=2.6700\,$d$^{-1}$. A phase-folded
grey-scale plot of the LSD profiles reveals variability with this frequency for
the HERMES data (left panel of Fig.\,\ref{LSDphasefold}, binned in 0.05 phase
intervals) but not for the HARPS data. This is { interpreted in terms of
complex multiperiodicity and mode beating that changes
drastically over time scales of weeks, as revealed by the CoRoT light
curve.} Keeping in mind that the spectroscopy suffers heavily from daily
aliasing, $f_{S1}$ { is equal to an alias of the CoRoT frequency $f_{31}$.}

As an additional test, we phase folded the residual LSD profiles with each of
the frequencies detected in the CoRoT photometry, { as well as with random
  frequencies}.  In that way, we found { only one} weak
coherent periodic signal in the HARPS LSDs for $f_3$ from visual inspection
(right panel of Fig.\,\ref{LSDphasefold}). Comparing both panels of
Fig.\,\ref{LSDphasefold}, we conclude that 
HD\,170580 has low-degree prograde $(f_{S1})$ and retrograde $(f_3)$
modes.

\subsection{Mode identification}
We attempted to identify the degree and azimuthal order of the two
dominant frequencies. Irrespective of the identification, this led to
$v\sin\!i=4\pm 2\,$km\,s$^{-1}$, where the uncertainty is dominated by the
various pulsational line-profile broadening levels depending on the kind of mode
that was assumed for $f_{S1}$ or $f_3$. Pulsational line-profile broadening was
so far ignored in the literature in deriving the rotational line broadening.

The HARPS data did not lead to any conclusive result on the identification for
$f_3$ due to the too low signal at the epoch of these spectra.  The HERMES data,
when fitted as if the star was a monoperiodic pulsator with a single mode
$f_{S1}$, led to a formal best line-profile fit for a prograde sectoral mode,
either with $l=+m=3$, $v\sin\!i\simeq 6\,$km\,s$^{-1}$, along with
$i\simeq 6^\circ$, where the uncertainty for the inclination is
$\sim\!3^\circ$. Given that we treated the star as monoperiodic, that the
dominant modes in space photometry usually have $l<3$, and that this solution is
close to an Inclination Angle of Complete Cancellation \citep[IACC, Table\,B1
in][]{Aerts2010}, we consider this result unlikely.  Indeed, it is a known
artifact of spectroscopic mode identification to find $l=3$ close to an IACC in
the case of low signal in the line-profile variability \citep[][for a thorough
discussion]{Chadid2001}.  Also, taking $v\sin\!i=4\pm 2\,$km\,s$^{-1}$ and
$i=6^\circ\pm 3^\circ$ leads to an equatorial rotation velocity that is
incompatible with the estimation of the rotation period from the detected
multiplets in the CoRoT data.  Moreover, as already mentioned, the detection of
zonal $(m=0)$, tesseral $(0\neq m\neq l)$, and sectoral ($(l=|m|)$) modes in the
space photometry as found from the multiplets indicated in
Table\,\ref{freqtable} and shown in Fig.\,\ref{windows} rather suggests a view
upon an intermediate inclination angle. Finally, the occurrence of helium
surface depletion can only be explained by atomic diffusion in action and this 
suggests slow rotation.

We conclude to have found marginal evidence of a low-degree prograde $(f_{S1})$
and retrograde $(f_3)$ mode in the spectroscopic line-profile variability of
HD\,170580 taken at two different epochs. As a consequence of the low
pulsational velocity amplitude, unambiguous identification of the two mode
degrees or azimuthal orders was not possible, despite the high quality of the
data. This also implies that the inclination angle could not be estimated properly.

\section{Comparison with stellar models}

Just as for the case of the CoRoT SPB target HD\,43317 \citep{Buysschaert2018},
we do not have a secured period spacing pattern of gravity modes at our disposal
to guide the asteroseismic modelling of HD\,170580. This prevented us from mode
identification via the method by \citet{VanReeth2016}, which was successfully
applied to { SPBs observed by {\it Kepler\/} with a ten times
better frequency resolution than the one of CoRoT long runs
\citep{Papics2017}}. We do have a secure identification of the mode degree $l$
from { six} rotationally-split multiplets.

In order to have an unrestricted range in mass for the search for stellar models
that can explain HD\,170580's asteroseismic properties listed in
Table\,\ref{freqtable}, we use the combined Particle Filter and Deep Neural
Network method (PF-DNN hereafter) developed by \citet{Hendriks2019}. This PF-DNN
is trained on an extensive grid of 352\,800 stellar models and their pulsation
frequencies of { radial ($l=0$), and zonal ($m=0$) dipole ($l=1)$ and
  quadrupole ($l=2$) modes.} The input physics of the models is fully described
in \citet{SchmidAerts2016}, to which we refer for details.\footnote{The SpaceInn
  grid of stellar models and their asteroseismic properties, as well as a script
  to run the PF-DNN are publicly available from {\tt
    fys.kuleuven.be/ster/Software/grid-massive-stars}} The PF-DNN allows fast
screening of millions of zonal ($m=0$) pulsation modes for non-rotating stellar
models representing stars in the core hydrogen burning phase and covering birth
masses between 2 and 20\,M$_\odot$. This PF-DNN screening has the aim to obtain
rough estimates of the six input parameters of the models: mass ($M$), initial
hydrogen mass fraction ($X$), metallicity ($Z$), central hydrogen mass fraction
($X_c$, a proxy of the stellar age), and two parameters connected with mixing of
the chemical elements beyond the Schwarzschild convective core boundary. These
two mixing parameters represent 1) convective core overshooting assuming an
exponentially decaying shape and overshoot distance $f_{\rm ov}$ using the
formulation by \citet{Freytag1996} and adopting the radiative temperature
gradient in the overshoot zone; 2) a constant level of diffusive mixing,
$D_{\rm mix}$, throughout the stellar envelope. The latter was found to be a
necessary ingredient in order to explain the mode trapping detected in {\it
  Kepler\/} photometry of gravity-mode pulsators of spectral type B
\citep{Moravveji2015,Moravveji2016}. { The models do not include atomic
  diffusion, because it is impossible to compute the appropriate radiative
  levitation for such an extensive grid. As shown by \citet{Deal2016} and by
  \citet{Aerts2018}, ignoring atomic diffusion for A-type and B-type stars can
  have major effects for the envelope mixing and hence for the pressure-mode
  frequencies, with deviations for the latter as large as 1\,d$^{-1}$.
The PF-DNN is therefore not
  expected to lead to an optimal solution for HD\,170580 if the measured
  frequencies listed in Table\,\ref{freqtable} involve pressure modes, but it
  might be suitable when involving only high-order gravity modes. In any case,
  application of the PF-DNN will point to the most appropriate stellar
  parameters within the limitations of the adopted input physics that went into
  the SpaceInn grid, which is state-of-the-art for seismic modelling of B-type
  pulsators following \citet{Moravveji2016}.}
\begin{figure*}[h!]
\centering
\includegraphics[width=0.95\linewidth]{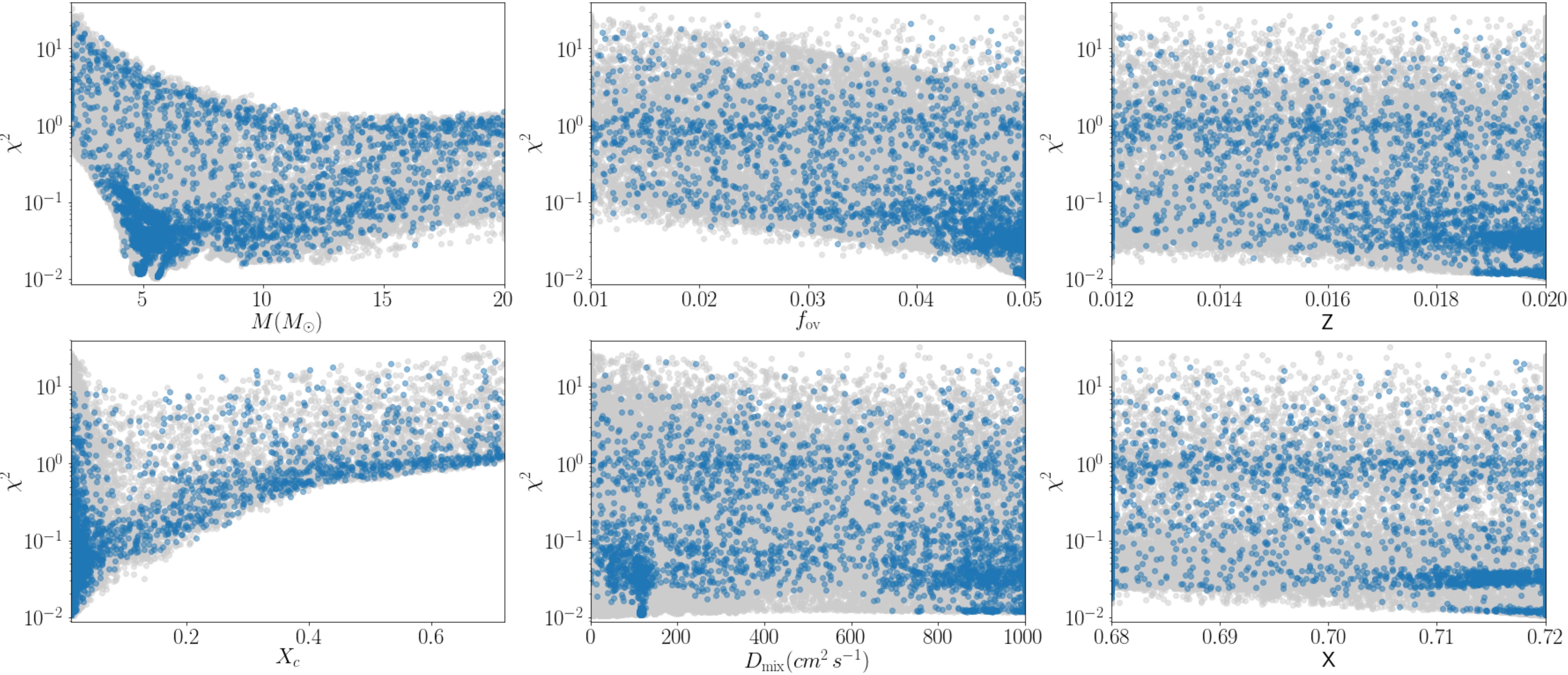}
\includegraphics[width=0.95\linewidth]{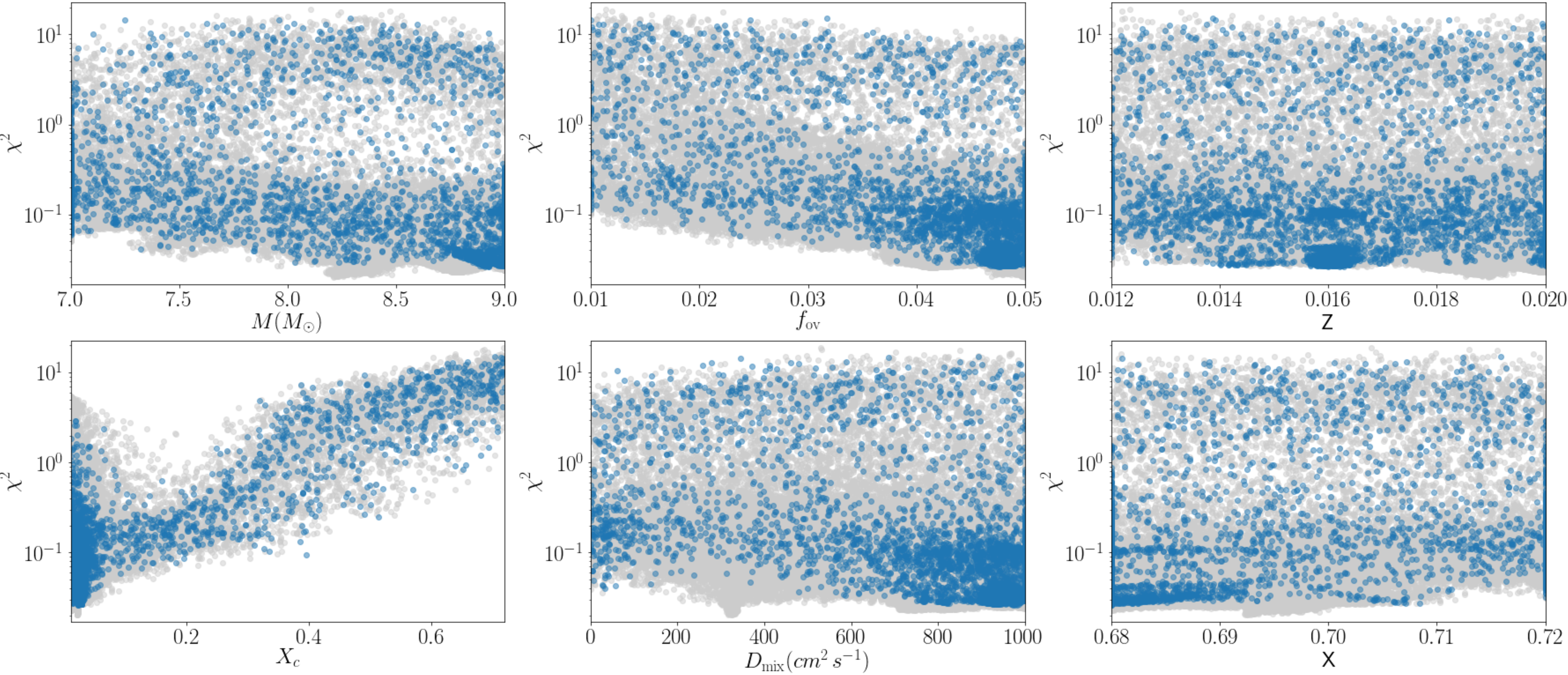}
\caption{Upper 6 panels: PF-DNN application to HD\,170580, based on 6
  frequencies ($f_1$, $f_2$, $f_{19}$, $f_{23}$, $f_{24}$ and $f_{26}$) of
  identified modes and 5 ($f_{10}$, $f_{14}$, $f_{16}$, $f_{27}$, and $f_{32}$)
  of unidentified modes.  Lower 6 panels: similar, but where the PF-DNN was
  retrained such as to comply with the Gaia DR2 parallax and the spectroscopic
  $T_{\rm eff}$, both at $2\sigma$-level.  The blue and grey points represent
  the results for two different settings for the genetic algorithm and sampling
  of the PF-DNN, according to \citet{Hendriks2019}.}
\label{DNN-Fig}
\end{figure*}

The PF-DNN searches for the optimal combination of the six free parameters on the
basis of the theoretically computed { zonal} pulsation frequencies of all the stellar
models in the grid, by comparing them with the measured pulsation frequencies of
the star by means of a $\chi^2$ distance measure, in the situation where limited
(or no) information on the mode identification of the frequencies is available.
The performance of the PF-DNN was tested on several well-known pulsators of
spectral type B that had been modelled previously and for which identification
of the degree $l$, azimuthal order $m$, and radial order $n$ of the
measured frequencies was already known from previous studies. This allowed
\citet{Hendriks2019} to assess the inference capacity of the PF-DNN in the
case where information on the { mode degrees $l$ was (partially) missing}. 
It was found that the PF-DNN performs optimally when the mode degrees
are known, while it becomes less performant as more of the degrees
are missing, with no inference capacity when none or only one of them is
known for a set of measured frequencies. { The PF-DNN is only suitable for
  slow rotators, as is the case for HD\,170580}.

We first applied the PF-DNN in the version of \citet{Hendriks2019} to
HD\,170580, relying only on the information derived from the CoRoT data, i.e.,
with { six zonal mode frequencies having an identification for the degree:
  $f_2$, $f_{19}$, $f_{23}$, $f_{24}$ and $f_{26}$ with $l=1$ and $f_1$ with
  $l=2$.  For these we thus search for an exact frequency match. The other five
  independent oscillation mode frequencies we used, without information on the
  mode degree, were $f_{10}$, $f_{14}$, $f_{16}$, $f_{27}$, and $f_{32}$. These
  are assumed to correspond with modes of $l\leq 2$.  From the measured
  rotational splittings of the six multiplets, we anticipate deviations from the
  zonal mode frequencies computed from the SpaceInn models to be zero 
for radial
  modes, in the range $[0.0;0.015\cdot\,(1-C_{nl})]$\,d$^{-1}$ for sectoral
  ($l=m$) dipole modes and up to twice this interval for sectoral quadrupole
  modes.}
\begin{table}
\tabcolsep=2pt
\centering
\caption{Best PF-DNN fit point for the densest sampling represented by the 
  grey dots in Fig.\,\ref{DNN-Fig}   for HD\,170580, in the case of 
  modelling based on just the CoRoT data 
  (left column) and from the CoRoT data in addition to luminosity
  constraints from Gaia DR2 and 
  the spectroscopic $T_{\rm eff}$, taking $2\sigma$ intervals for those
  two  quantities (right column).}
\label{DNN-Table}
\begin{tabular}{l|c|c}
Parameter & Only CoRoT & CoRoT, Gaia DR2, $T_{\rm eff}$ \\
\hline
$M$ (M$_\odot$)    & 5.6  & 8.2      \\
$f_{\rm ov}$ ($H_p$)   & 0.05      & 0.05   \\
$Z$                          & 0.020       & 0.019     \\
$X_c$       & 0.010       & 0.021    \\
$D_{\rm mix}$\!(cm$^2$\,s$^{-1}$)  & 1.3        & 324   \\
$X$                          & 0.72       & 0.69          \\
\hline
r.m.s.\ $f$ (d$^{-1}$) & 0.010 & 0.020 \\
\hline
\hline
Freq. $(d^{-1})$ & \multicolumn{2}{c}{Mode Identification ($m=0$)} \\
\hline
$f_{27}=3.7348$ & $l=2, n=0$ & $l=2, n=0$ \\
$f_{23}=3.6568$ & $l=1$(fixed), $n=+2$ & $l=1$(fixed), $n=+2$ \\
$f_{14}=3.4432$ & $l=2, n=-1$ & $l=2, n=-1$ \\
$f_1=3.3373$ & $l=2$(fixed), $n=-2$ & $l=2$(fixed), $n=-2$ \\
$f_2=3.2371$ & $l=1$(fixed), $n=+1$ & $l=1$(fixed), $n=+1$ \\
$f_{32}=2.2010$ & $l=1, n=-3$ & $l=1, n=-1$ \\
$f_{10}=1.9691$ & $l=1, n=-5$ & $l=2, n=-6$ \\
$f_{24}=1.8495$ & $l=1$(fixed), $n=-6$ & $l=1$(fixed), $n=-2$ \\
$f_{16}=1.8096$ & $l=1, n=-7$ & $l=2, n=-7$ \\
$f_{26}=1.7849$ & $l=1$(fixed), $n=-8$ & $l=1$(fixed), $n=-3$ \\
$f_{19}=1.7083$ & $l=1$(fixed), $n=-9$ & $l=1$(fixed), $n=-4$ \\
\end{tabular}
\end{table}

The results of the PF-DNN application for the frequency matching based on the
{ 11} frequencies (indicated as red dashed lines in Fig.\,\ref{freqspectrum})
are shown in the upper six panels of Fig.\,\ref{DNN-Fig}.  The blue points are
the results for the basic setting of the genetic algorithm and sampling of the
PF-DNN, while the grey points represent a more dense setting that samples more
points per iteration (see \citet{Hendriks2019} for details).  It can be seen
{ from the global trends in Fig.\,\ref{DNN-Fig}} that the metallicity,
initial hydrogen fraction, and level of particle mixing in the stellar envelope
are hardly constrained, but the mass, core overshooting, and evolutionary stage
can be estimated.  { This is a similar result than for two previous
  asteroseismic modelling applications to CoRoT B pulsators
  \citep{Degroote2010,Buysschaert2018}.  The best solution found by the PF-DNN
  with densest (grey dots) sampling is listed in the left column of
  Table\,\ref{DNN-Table} and involves low-order modes with $n\in [-9,+2]$,
  leading to an r.m.s.\ of 0.010\,d$^{-1}$.}

\begin{figure*}
\centering
\rotatebox{270}{\resizebox{7cm}{!}{\includegraphics{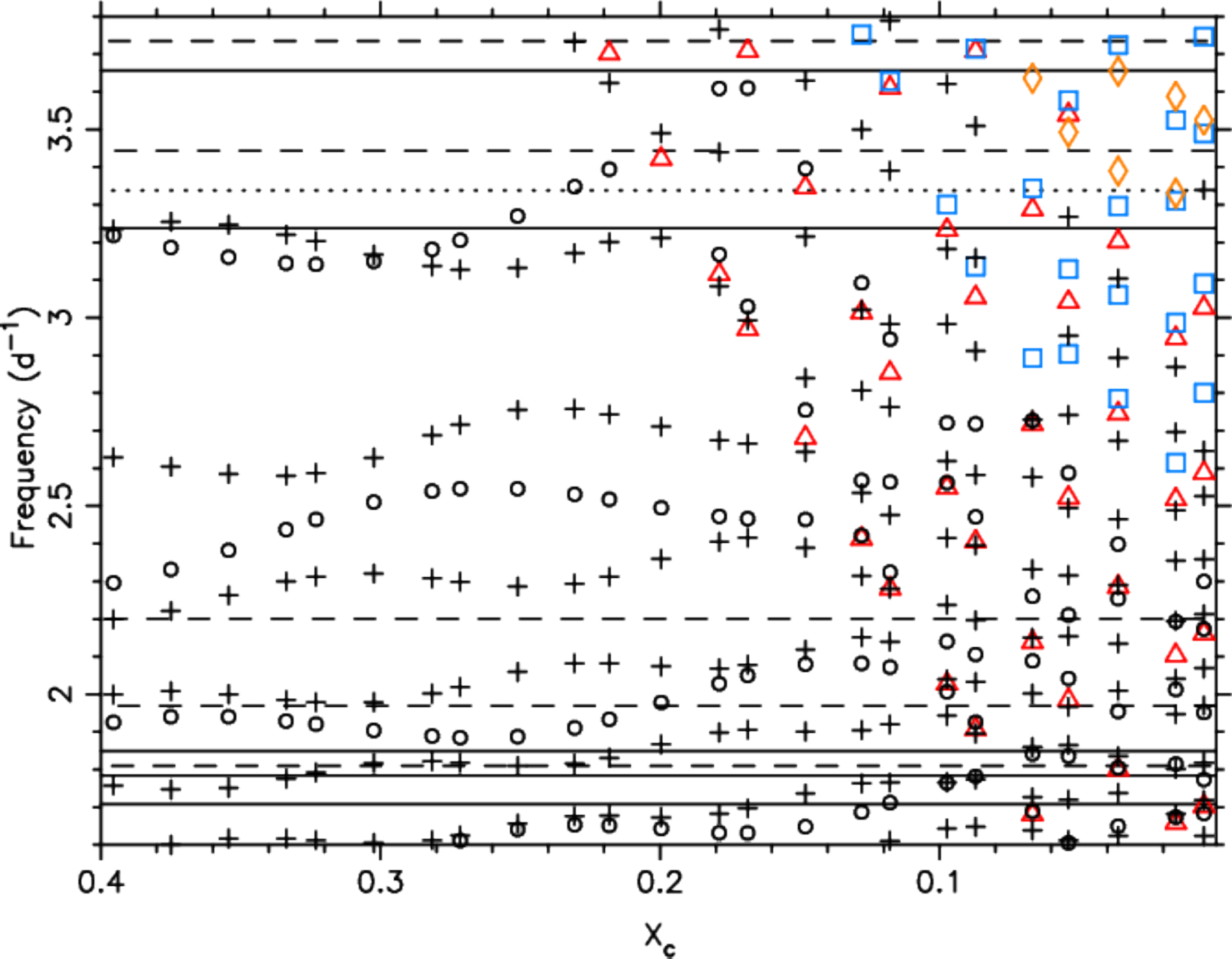}}}
\rotatebox{270}{\resizebox{7cm}{!}{\includegraphics{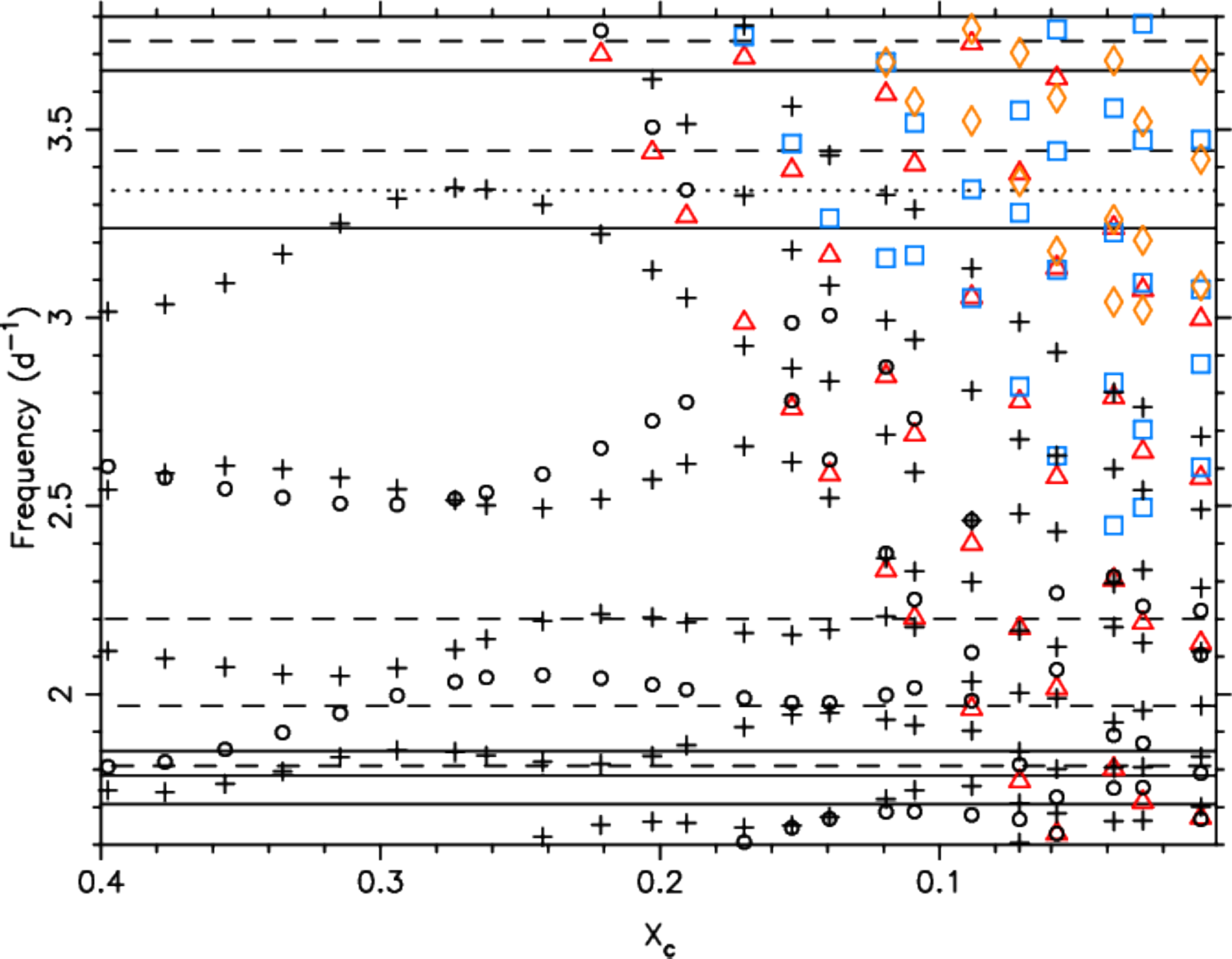}}}
\rotatebox{270}{\resizebox{7cm}{!}{\includegraphics{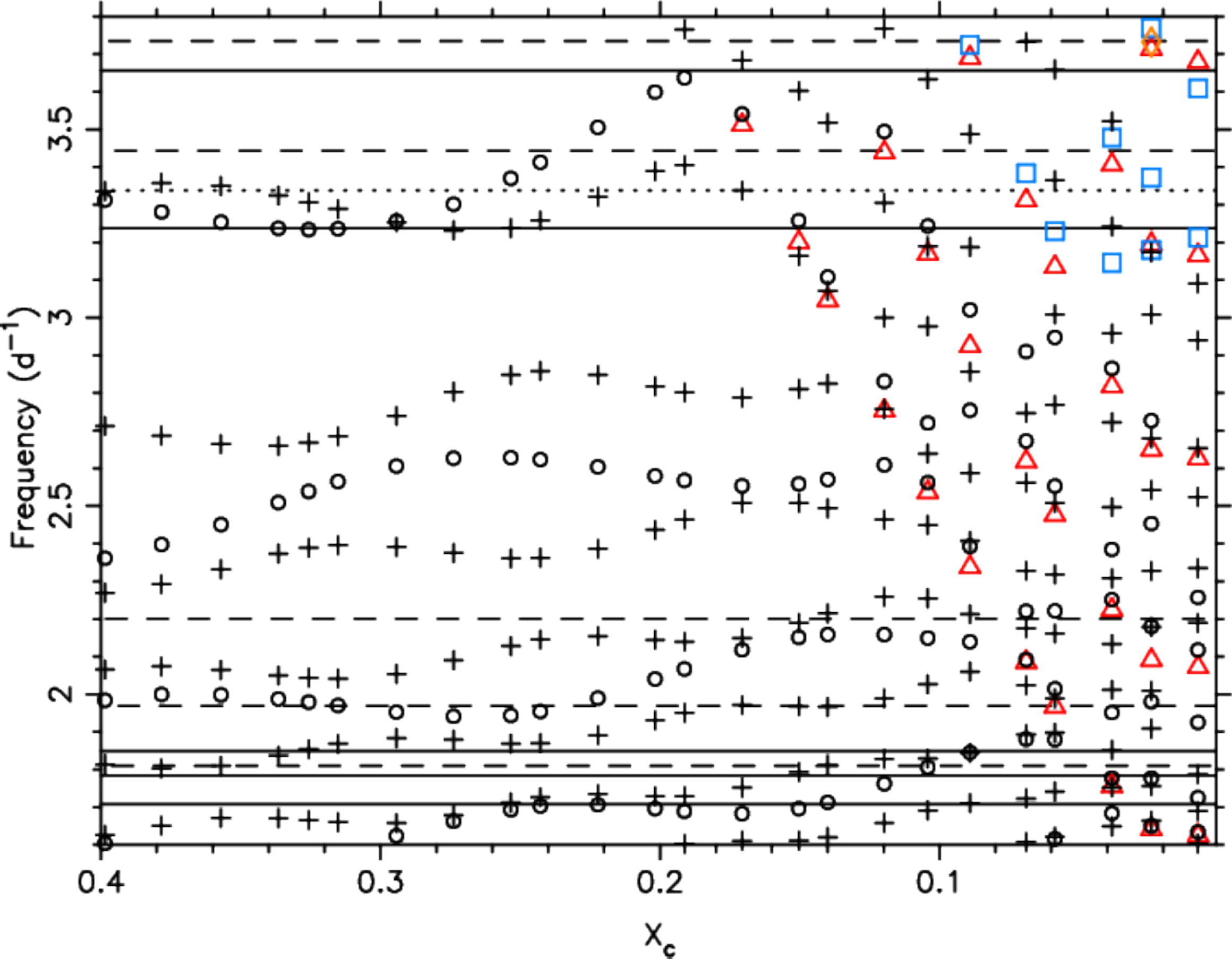}}}
\rotatebox{270}{\resizebox{7cm}{!}{\includegraphics{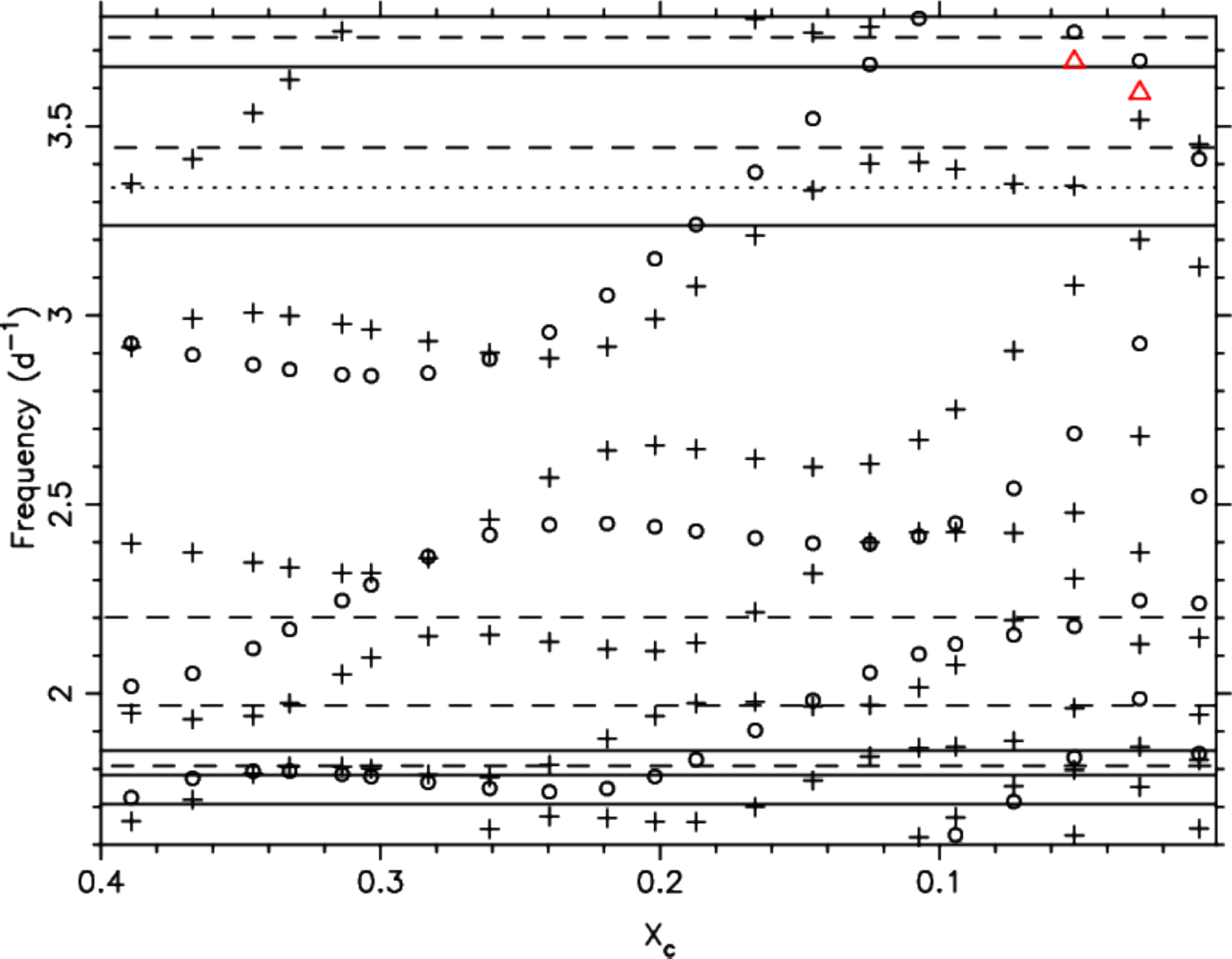}}}
\caption{{Evolution of the zonal mode frequencies of stellar models with
  input parameters 
$M=6\,$M$_\odot$, $X=0.72$, $Z=0.020$, $f_{\rm ov}=0.05$, and
  $D_{\rm mix}=1\,$cm$^2$\,s$^{-1}$ (upper left), $M=8\,$M$_\odot$, $X=0.70$,
  $Z=0.020$, $f_{\rm ov}=0.05$, and $D_{\rm mix}=100\,$cm$^2$\,s$^{-1}$ (upper
  right), $M=6\,$M$_\odot$, $X=0.72$, $Z=0.012$, $f_{\rm ov}=0.05$, and
  $D_{\rm mix}=1\,$cm$^2$\,s$^{-1}$ (bottom left), $M=8\,$M$_\odot$, $X=0.70$,
  $Z=0.020$, $f_{\rm ov}=0.01$, and $D_{\rm mix}=100\,$cm$^2$\,s$^{-1}$ (bottom
  right).  The panels show the zonal mode frequencies in the region of the
  measured frequencies of HD\,170580, which are indicated as horizontal
  lines (dotted: $l=2$, full: $l=1$, dashed: unknown $l$). The model frequencies
  are code as black circles: $l=1, n<0$, black plus signs: $l=2, n<0$, red
  triangles: $l=0$, blue squares: $l=1, n\geq 0$, orange diamonds:
  $l=2, n\geq 0$. }}
\label{mode-diagram}
\end{figure*}

By construction, the best PF-DNN results come from an interpolation of the modes
belonging to the ``true'' stellar models in the SpaceInn grid.  { The
  evolution of the zonal-mode frequencies of the models closest in mass to this
  solution in the SpaceInn grid is shown in the upper left panel of
  Fig.\,\ref{mode-diagram} for the second half of the main sequence.  The mode
  identification in Table\,\ref{DNN-Table} and estimated $X_c$ correspond with
  $(1-C_{nl})\in [0.51; 0.60]$ for the $l=1$ modes and
  $(1-C_{nl})\in [0.86; 0.97]$ for the $l=2$ modes. This implies a frequency
  shift for the $l=2, n=0$ and $l=2, n=-1$ modes of $\simeq 0.024\,$d$^{-1}$
  based on the measured splitting of the $l=2, n=-2$ mode in the case of rigid
  rotation.  The symbols sizes of all the modes in Fig.\,\ref{mode-diagram} have
  been chosen to cover such a typical frequency shift.  The found r.m.s.\ of
  0.010\,d$^{-1}$ for the difference between the 11 measured and model
  frequencies points to an overall good frequency match given the unknown
  azimuthal order for five of the 11 modes and their anticipated frequency
  shifts due to Ledoux splitting.  

  Close inspection of the upper left panel of Fig.\,\ref{mode-diagram} shows
  that the fit is only good for the $l=1, n<-2$ modes and only for models at a
  very advanced stage of evolution where the mode density is large and mode
  bumping occurs \citep[e.g.,][]{Shibahashi1976,Aizenman1977}.  
Moreover, modes
  of evolved models with high density contrast close to the TAMS may also be
  subject to the phenomenon of avoided crossings \citep[e.g.,][for adiabatic and
  non-adiabatic computations, respectively]{Gabriel1980,JCD1981}.  Both these
  phenomena are visible in the top right corners of the panels in
  Fig.\,\ref{mode-diagram}, where the smooth evolution of the modes deteriorates
  for $X_c<0.1$.  We refer to \citet[][Chapter\,11]{Smeyers2010} for a thorough
  mathematical description and concrete numerical examples of 
mode bumping and
  avoided crossings, but point out here that these may induce appreciable
  frequency shifts, much larger than the measured frequency uncertainties from
  the CoRoT data.  Moreover, as already discussed above, deviations for the
  pressure modes are anticipated from the lack of atomic diffusion in the
  models.  Hence, both at the level of the equilibrium models (atomic diffusion)
  and and the level of the pulsation computations (mode bumping and avoided
  crossings), we face theoretical uncertainties with possible considerable
  frequency shifts in near-TAMS stars. This typically occurs for the pressure,
  gravity, and fundamental non-radial modes of low radial orders, as we are
  dealing with here.  }

Confrontation of the result of this ``CoRoT-only'' modelling in the left column
of Table\,\ref{DNN-Table} and the upper left panel of Fig.\,\ref{mode-diagram}
leads to a stellar model that has $\log(L/L_\odot)\simeq 3.55$ and
$T_{\rm eff}\simeq 11\,700\,$K, which are not compliant with the Gaia DR2
luminosity we derived nor with the spectroscopic $T_{\rm eff}$.  { Moreover,
  HD\,170580 has a low surface metallicity of $Z=0.0095\pm 0.0030$.  In order to
  assess the effect of the metallicity, we show in the bottom left panel of
  Fig.\,\ref{mode-diagram} the modes for a model with the same parameters than
  the upper panel except that $Z=0.012$, which is the lowest metallicity in the
  SpaceInn grid.  This gives a worse fit, because the mode frequencies get
  shifted to higher values and there are no $l=2, n\geq 0$ modes available in
  the measured frequency range.}

{ A decrease in $Z$ is degenerate with an increase in mass when fitting gravity
  modes \citep[][Fig.\,5]{Moravveji2015}. Since a higher
  luminosity is also needed to meet the Gaia DR2 luminosity, } we 
performed
``constrained'' modelling, demanding compliance with both the Gaia DR2
luminosity and the spectroscopic $T_{\rm eff}$, taking $2\sigma$ level for
both. We thus retrained the PF-DNN on the basis of stellar models in the
SpaceInn grid that are situated within the box in Fig.\,\ref{HRD}.  This does
not imply that the PF-DNN will deliver a best solution within that error box,
but rather that it is trained only on the basis of models within that box. It
might extrapolate outside the box in Fig.\,\ref{HRD} to find the best
solutions.  Application of this version of the PF-DNN results in the patterns
shown in the lower 6 panels in Fig.\,\ref{DNN-Fig} and the best solution listed
in the right column of Table\,\ref{DNN-Table}.  It is seen from
Fig.\,\ref{DNN-Fig} that the restriction imposed by the Gaia astrometry and the
spectroscopy in the training of the PF-DNN does not change the overall
morphology in the $\chi^2$ space in terms of evolutionary stage or level of core
overshooting. { The best estimate for the stellar mass now occurs at
  $M=8.2\,$M$_\odot$ for a model with $\log(L/L_\odot)\simeq 4.0$. This model
  identifies the radial orders from $n=-7$ to +2 and has an r.m.s.\ for the
  frequencies of 0.020\,d$^{-1}$, i.e., twice as large as for the unconstrained
  ``CoRoT-only'' modelling but still fine given the expected Ledoux shifts for
  the five frequencies with unidentified degree.  Also here the pressure modes
  are the ones that are not well reproduced (upper right panel of
  Fig.\,\ref{mode-diagram}).  Moreover, this model has a too high luminosity
  and a too low $T_{\rm eff}$. In comparison, we show in the bottom right
  panel of Fig.\,\ref{mode-diagram} a model for the same parameters, except that
  the core overshooting is only $f_{\rm ov}=0.01$ instead of 0.05. This model
  does occur at the edge of the $2\sigma$ 
Gaia DR2 error box in Fig.\,\ref{HRD} but it
  lacks pressure modes in the measured frequency regime. }

{ Because some of the lower-amplitude multiplets 
have components that cannot
  be distinguished from combination frequencies, we finally applied the PF-DNN
  for the case of only having a fixed mode degree for the quintuplet and the two
  dominant triplets (with zonal frequencies $f_1, f_2$, and $f_{19}$). This led
  to a near-TAMS model with $M=9.8\,$M$_\odot$, $f_{\rm ov} = 0.045$ for
  ``CoRoT-only'' modelling. That model has an r.m.s.\ of 0.009\,d$^{-1}$ but
  $\log(L/L_\odot)\simeq 4.31$, which is not compliant with the Gaia
  luminosity. It also has a far too low $T_{\rm eff}$. The constrained modelling
  version in this case led to a model with $M=8.6\,$M$_\odot$ and similar
  parameters than those in the right column of Table\,\ref{DNN-Table}, for an
  r.m.s.\ of 0.017\,d$^{-1}$ and a similar deficit of appropriate pressure-mode
  frequencies.  From this we conclude there to be similar results if we feed the
  PF-DNN with only three instead of six identified zonal modes.}

We conclude from combined CoRoT, Gaia DR2 and spectroscopic data that HD\,170580
is a star of $M\simeq 8\,$M$_\odot$ close to the TAMS, but that stellar models
without atomic diffusion cannot explain its detected { pressure-mode}
frequencies to within the measurement errors.  { The Ledoux coefficients for
  the three dipole gravity 
modes with the lowest frequency and highest 
radial order in the model closest to the
  parameters listed in the right column of Table\,\ref{DNN-Table}
  $(f_{24},f_{26},f_{19})$ give $(1-C_{n1})=0.6212, 0.8202, 0.5424$ for
  $n=-2, -3, -4$, respectively.  The measured splittings for these three modes
  then lead to a rotation frequency of 
0.0208$\pm$0.0017, 0.0140$\pm$0.0020, and
  0.0250$\pm$0.0016\,d$^{-1}$.  This is a tentative indication that the star
  might have mild differential rotation in its interior. However, given that we
  are in an asteroseismic regime of low-order modes where the Ledoux
  splitting is model-dependent and we rely on models incompatible with the Gaia
  luminosity in terms of their pressure modes, we consider this a preliminary
  result at best. Stellar models with atomic diffusion that are compatible with all the
  data (CoRoT, Gaia, spectroscopy) need to be constructed to evaluate this
  suggestive result further.}

\section{Discussion and future improvements}

We detected { 42} oscillation mode frequencies in the CoRoT light curve of
HD\,170580 and selected { 11} independent mode frequencies for forward
asteroseismic modelling.  Among those { 11}, { five zonal dipole modes and
  one quadrupole mode} occur. The other { 5} frequencies were fitted while
allowing $l=0,1,2$ for their degree. { Despite huge efforts to gather
  extensive time-series spectroscopy with some the best spectrographs in the
  world, that data did not add observational constraints to the modelling
  compared to earlier snapshot spectroscopy as only a better estimate of
  $v\sin i$ and its error was achieved.  Intensive time-series spectroscopy
  for asteroseismology should better be focused on pulsating spectroscopic
  binaries for tidal asteroseismology, where there is a lack of
high-precision data at a level suitable to improve their stellar interior models. 
For single stars, spectroscopic monitoring is better
  done after the analysis of the space photometry and only for stars with
  mmag-type dominant modes that lack an identification of their $(l,m)$.}

{ The asteroseismic modelling of HD\,170580} was done from a Particle Filter
combined with a Deep Neural Network trained on the basis of a large grid of
stellar models covering the mass range $[2,20]\,$M$_\odot$, as well as a
retrained version limited to the luminosity and effective temperature range from
Gaia and spectroscopy, respectively.  We found good agreement with the CoRoT
data to within the frequency precision { of the detected low-order gravity
  modes} for a near-TAMS star of $M\simeq 8\,$M$_\odot$ with a large core
overshooting.  However, the models with these parameters are not compliant with
the { low-order pressure modes and only marginally so with the Gaia data.}
The latter did not treat the poorly known multiplicity of HD\,170580 yet. The
asteroseismic models lack atomic diffusion for computational reasons, but the
spectroscopy of HD\,170580 { reveals} that it is active in the star.  {
  Moreover, we limited to adiabatic pulsation computations.  This lack of atomic
  diffusion and/or imperfect treatment of the phenomena of mode bumping and
  avoided crossings} could be reasons why the combined asteroseismic,
astrometric, and spectroscopic modelling of HD\,170580 cannot explain the CoRoT
frequencies { of the low-order pressure modes} to within their measured precision.

HD\,170580 revealed itself as an interesting asteroseismic target, given that it
is close to the TAMS. This holds the potential to achieve a high-precision
asteroseismic measurement of the amount of helium in its core, just prior to the
start of its blue supergiant phase.  { Moreover, we found a hint of a low
  level of differential rotation from its low-order gravity modes}.  However,
before we can deduce { firm asteroseismic conclusions on these two aspects,
  improvement in the modelling is required.} This may come from three fronts:
better oscillation mode frequency precision if TESS would observe the star, good
predictions of low-order non-radial oscillation modes from stellar models
with atomic diffusion including radiative levitation, and a Gaia DR3 parallax
taking proper account of the multiplicity of the star.

\begin{acknowledgements}
  This project has received funding from the European Research Council (ERC)
  under the European Union's Horizon 2020 research and innovation programme
  (grant agreement N$^\circ$670519: MAMSIE) and was supported by the National
  Science Foundation of the United States under Grant N$^\circ$NSF PHY--1748958.
  The computation of the stellar model grid used was funded by the European
  Community's Seventh Framework Programme FP7-SPACE-2011-1, project number
  312844 (SPACEINN).  A.T. is supported as a Research Associate at the Belgian
  Scientific Research Fund (F.R.S-FNRS).  C.A. thanks Dr.\,Eric Michel for the
  memorable efforts in trying to get the best stars selected as primary targets
  in CoRoT's asteroseismology CCDs.  We are grateful to all observers from the
  Institute of Astronomy of KU\,Leuven who contributed to the gathering of the
  HERMES spectroscopy at the Mercator telescope and to the MESA/GYRE developers
  for their continuous efforts and open access policy for their codes.
\end{acknowledgements}

\bibliographystyle{aa}
\bibliography{AA201834762R}

\end{document}